\documentclass[10pt, conference, letterpaper]{IEEEtran}
\IEEEoverridecommandlockouts
\usepackage{cite}
\usepackage{amsmath,amssymb,amsfonts}
\usepackage[noend]{algorithmic}

\usepackage{graphicx}
\usepackage{float}

\usepackage{multirow}

\usepackage{textcomp}
\usepackage{xcolor}
\def\BibTeX{{\rm B\kern-.05em{\sc i\kern-.025em b}\kern-.08em
    T\kern-.1667em\lower.7ex\hbox{E}\kern-.125emX}}

\usepackage{float}
\usepackage{algorithm}
\usepackage{subfigure}
\usepackage{makecell}
\usepackage{wrapfig}
\usepackage{rotating}

\usepackage{xcolor}

\usepackage{footmisc}

\newtheorem{definition}{Definition}
\newtheorem{theorem}{Theorem}
\newtheorem{lemma}{Lemma}
\DeclareMathOperator*{\argmax}{arg\,max}

\makeatletter
\newcommand{\removelatexerror}{\let\@latex@error\@gobble}
\makeatother

\usepackage{fancyhdr}

\begin{document}
	\bstctlcite{IEEEexample:BSTcontrol}

\title{PrivShape: Extracting Shapes in Time Series under User-Level Local Differential Privacy
}

\author{
\IEEEauthorblockN{Yulian Mao\textsuperscript{1,2}, Qingqing Ye\textsuperscript{2 *}, Haibo Hu\textsuperscript{2}, Qi Wang\textsuperscript{1,3}, Kai Huang\textsuperscript{4}}\\
\textsuperscript{1}{\small Department of Computer Science and  Engineering, Southern University of Science and Technology} \\

\textsuperscript{2}{\small Department of Electrical and Electronic Engineering, The Hong Kong Polytechnic University}\\
\textsuperscript{3}{\small Research Institute of Trustworthy Autonomous Systems, Southern University of Science and Technology}\\
\textsuperscript{4}{\small School of Computer Science and Engineering, Macau University of Science and Technology}\\
{\small yulian.mao@connect.polyu.hk, \{qqing.ye, haibo.hu\}@polyu.edu.hk,} {\small wangqi@sustech.edu.cn, kylehuangk@gmail.com}
}

\maketitle
\thispagestyle{fancy}
\fancyhead[C]{\textcolor{red}{This paper has been accepted by IEEE 40th Annual International Conference on Data Engineering (ICDE2024)}}

\begin{abstract}
Time series have numerous applications in finance, healthcare, IoT, and smart city. In many of these applications, time series typically contain personal data, so privacy infringement may occur if they are released directly to the public. Recently, local differential privacy (LDP) has emerged as the state-of-the-art approach to protecting data privacy. However, existing works on LDP-based collections cannot preserve the shape of time series. A recent work, PatternLDP, attempts to address this problem, but it can only protect a finite group of elements in a time series due to $\omega$-event level privacy guarantee. In this paper, we propose  PrivShape, a trie-based mechanism under user-level LDP to protect all elements. PrivShape first transforms a time series to reduce its length, and then adopts trie-expansion and two-level refinement to improve utility. By extensive experiments on real-world datasets, we demonstrate that PrivShape outperforms PatternLDP when adapted for offline use, and can effectively extract frequent shapes.
\end{abstract}


\vspace{-1em}
\footnote{* Dr. Qingqing Ye is the corresponding author.}

\section{Introduction}
Time series data are being generated on a large scale across a wide range of application domains, such as industrial IoT, finance, healthcare monitoring, operational event logs, and smart home sensors.
Extracting features from time series is usually the first step towards understanding and mining time series data. In particular, as a synopsis of a time series, its {\em shape} is a crucial feature that records the information of value change and trend. Shape is often measured by a distance metric~\cite{Paparrizos2015}. Many time series mining techniques are based on shape analysis \cite{Paparrizos2015, Siddiqui2020, Kowsar2022}. However, directly applying these techniques for shape analysis may cause serious privacy breaches, thus requiring privacy-preserving mechanisms.

To address privacy concerns when collecting data from users, local differential privacy (LDP) has been proposed~\cite{Kasiviswanathan2008, Cormode2018, Ye2020}. Under LDP, users  perturb their data locally before uploading it for further processing. The local perturbation is controlled by a parameter known as privacy budget, denoted by $\epsilon$. LDP ensures that even if an adversary obtains all the noisy data, the true value of a specific user still cannot be inferred with high confidence. For time series data, a line of work has studied the perturbation mechanisms in the context of LDP~\cite{Wang2020, Ye2021, Wang2021a, Ren2022}, which are classified into three categories in terms of their privacy levels. Specifically, {\em event-level privacy}~\cite{Dwork2010} guarantees the privacy of an individual element~\cite{Ye2021, Ren2022}, {\em $\omega$-event privacy}~\cite{Kellaris2014} protects a group of consecutive $\omega$ elements~\cite{Wang2020}, and {\em user-level privacy}~\cite{Dwork2010} protects the entire time series of a user~\cite{Ahuja2023, Dong2023}. Obviously, while {\em user-level privacy} offers the strongest privacy guarantee for individual data, achieving it becomes challenging in practice because of the heavy perturbation to ensure privacy. 

As of now, PatternLDP~\cite{Wang2020} is known as the only LDP mechanism that attempts to preserve shapes in time series data publishing. It extracts shape features by sampling trend-related elements in the time series within a $\omega$-length window and subsequently allocates privacy budgets among them to satisfy $\omega$-event privacy. However, when it comes to the stricter user-level privacy, the performance of PatternLDP deteriorates significantly as the entire time series is allocated a single privacy budget $\epsilon$. As such, the budget for each selected element becomes extremely small, resulting in heavy perturbation that significantly distorts the original shape. Furthermore, PatternLDP perturbs each time series individually and cannot exploit the shape information from time series belonging to users of the same type, which may exhibit common shape features. Thus, PatternLDP is not particularly effective for shape extraction under user-level privacy.

In this paper, we study more effective ways to extract essential shapes in time series under user-level LDP. Essential shapes are those shapes that outline the overall shape characteristics of a time series dataset, such as trends, patterns, or structures. Here, we use two simple examples to illustrate essential shapes in time series. 

\textbf{Example I:} (\textit{Motion Time Series}) The user's gesture can be captured as a time series of motion trajectory through sensor readings. When individuals perform the same gesture at different speeds, the silhouettes of the time series exhibit a similar shape but vary in the length of the recorded time series (e.g., a longer time series from an elderly). When the speed of the gesture is not a concern, we can shorten the time series by removing similar or even repeated values while extracting essential shapes to ascertain the specific changes in different gestures.

\begin{figure}[htbp]
	\vspace{-1em}
	\centering
	\subfigure[Features of  ``No" ]{
		\label{mfcc}
		\includegraphics[width=0.13\textwidth]{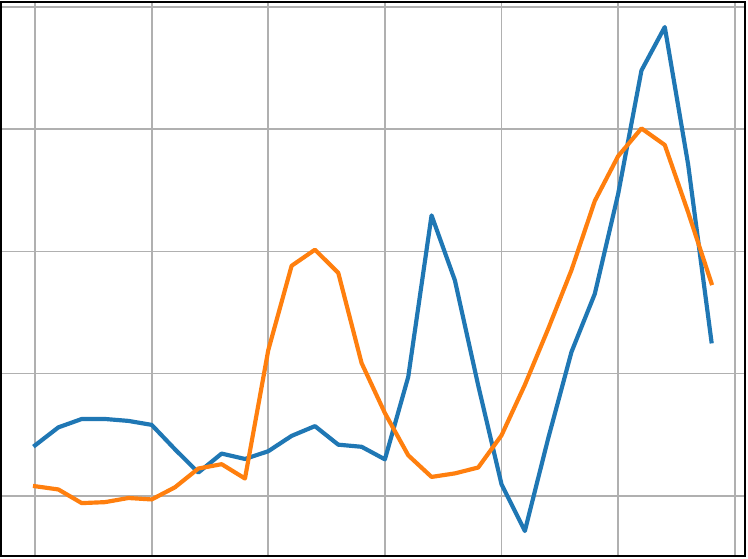}}\hspace{0.01\textwidth}
	\subfigure[ Essential shape]{
		\label{essential}
		\includegraphics[width=0.13\textwidth]{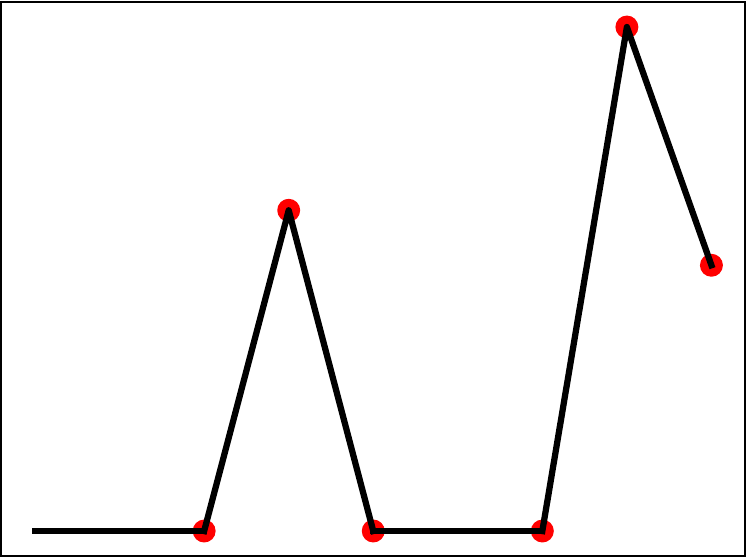}}\hspace{0.01\textwidth}
	\subfigure[After SAX]{
		\label{mfccsax}
		\includegraphics[width=0.15\textwidth]{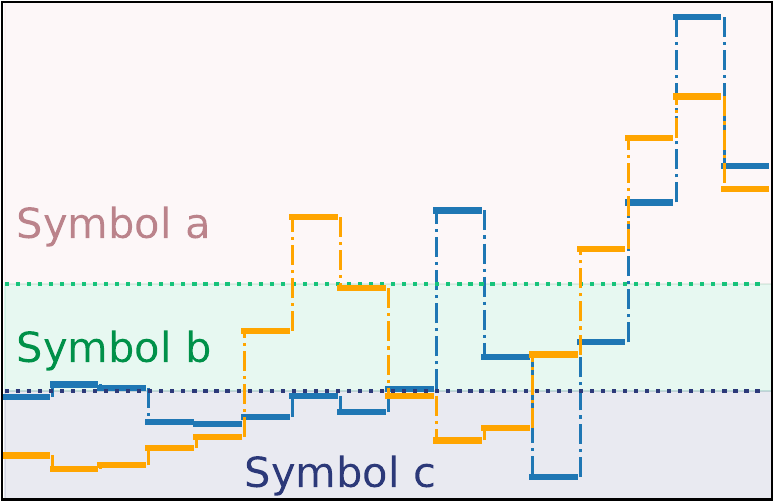}}
	\vspace{-0.5em}
	\caption{(a) illustrates the shapes of frequency features corresponding to the pronunciations of ``No" by two speakers. The two time series cannot be matched exactly, but they do have similar essential shapes (red points) shown in (b). (c) shows the essential shapes captured from the original time series in (a) by SAX.}			
	\label{frequency}
	\vspace{-0.5em}
\end{figure}

\textbf{Example II:} (\textit{Speech Time Series}) The user's speech acoustics can be represented as a time series. However, when different individuals pronounce the same content, the frequency features may vary in length due to differences in speech rate (see Fig.~\ref{mfcc}). Nevertheless, the frequency features of the same phoneme pronounced by different individuals often exhibit a similar shape, as referenced in \cite{Hamooni2014,Li2017a}. For instance, the two time series in Fig.~\ref{mfcc} can be compressed into the red points presented in Fig.~\ref{essential}. Although there are fewer data points, the essential shape well exhibits the trends and structures of the original time series. 

\begin{figure}[!htb]
	\vspace{-0.5em}
	\centering  
	\subfigure[Scaling.]{
		\label{scaling}
		\includegraphics[width=0.15\textwidth]{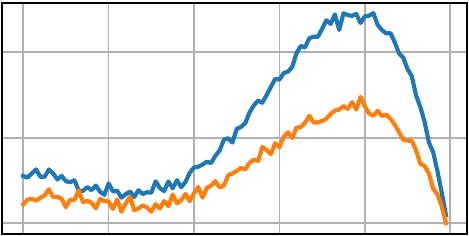}}\hspace{0.02\textwidth}
	\hspace{0.025\textwidth}
	\subfigure[Not warping.]{
		\label{warping}
		\includegraphics[width=0.15\textwidth]{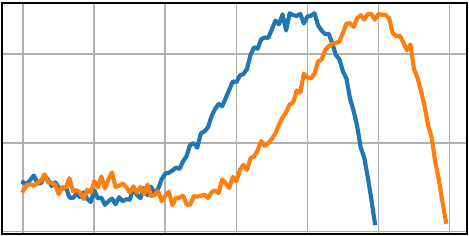}}
	\vspace{-0.7em}
	\caption{(a) illustrates that two time series exhibit scaling along value-axis despite having similar shapes. (b) elucidates that the lack of warping between the two time series is attributed to delays along the time-axis.}
	\label{timeseriesproblem}
	\vspace{-0.7em}
\end{figure}

Since essential shapes are frequent within the same class (e.g., the essential shape in Fig.~\ref{essential} of ``No"), extracting essential shapes among the entire dataset can be regarded as a frequent shape mining task. However, extracting shapes is challenging due to noise, scaling, and time not warping (see Fig.~\ref{timeseriesproblem} for explanation), which prevents exact matching of time series on the value-axis and time-axis. 

To address these challenges, we can compress the values on the value-axis by time series transformation techniques, such as Symbolic Aggregate approXimation (SAX)~\cite{Lin2007, Li2020, Liu2022}  (see Fig.~\ref{mfccsax} as an example).
Compressive SAX can further compress the time-axis to address the challenge of time not warping by removing repeated symbols while retaining symbols can still capture trend changes.


Such symbol removal in Compressive SAX is also beneficial for LDP perturbation, which reduces the number of data points to be perturbed. Leveraging this transformation, we introduce {\it PrivShape}, a user-level LDP mechanism for extracting time series shapes. Our mechanism generates a series of candidate shapes using a trie structure and perturbs a user's selection among these candidates instead of directly adding noise to users' sequences. Furthermore, we present a pruning strategy based on trie expansion and a two-level refinement strategy to enhance utility. Our contributions are summarized as follows.
%
\begin{itemize}
\item  We address the utility issue of shape extraction in time series by focusing on a sequence of essential shapes and propose a user-level LDP mechanism PrivShape. To the best of our knowledge, this is the first attempt to extract shapes in time series under user-level local differential privacy.
\item We design two optimization strategies for  PrivShape to enhance utility, namely trie-expansion pruning and two-level refinement, which allows for efficient utilization of the privacy budget.
\item To evaluate the effectiveness of  PrivShape, we conduct extensive experiments over two benchmark datasets. For the sake of fairness, we extend PatternLDP into user-level privacy and accommodate its privacy budget allocation strategy for offline use. The experimental results consistently demonstrate that PrivShape outperforms the existing mechanism significantly.
\end{itemize}

The remainder of this paper is organized as follows. Section~\ref{statement} introduces the preliminaries and formulates the problem. Section \ref{baseline} and Section \ref{improvement} propose the baseline mechanism and the enhanced mechanism PrivShape, respectively. Section~\ref{experiment}
presents the experimental results. Finally, we review the existing works in
Section~\ref{relatedwork} and conclude this paper in Section~\ref{conclusion}.

\vspace{0em}

\section{Preliminaries and Problem Formulation}\label{statement}
In this section, we first introduce the preliminaries on Symbolic Aggregate approXimation (SAX) and local differential privacy, and then present a formal problem statement.

\vspace{-0.5em}
\subsection{Symbolic Aggregate Approximation on Time Series}\label{sax_info}

\begin{figure}[!htb]
	\centering
	\includegraphics[width=0.45\textwidth]{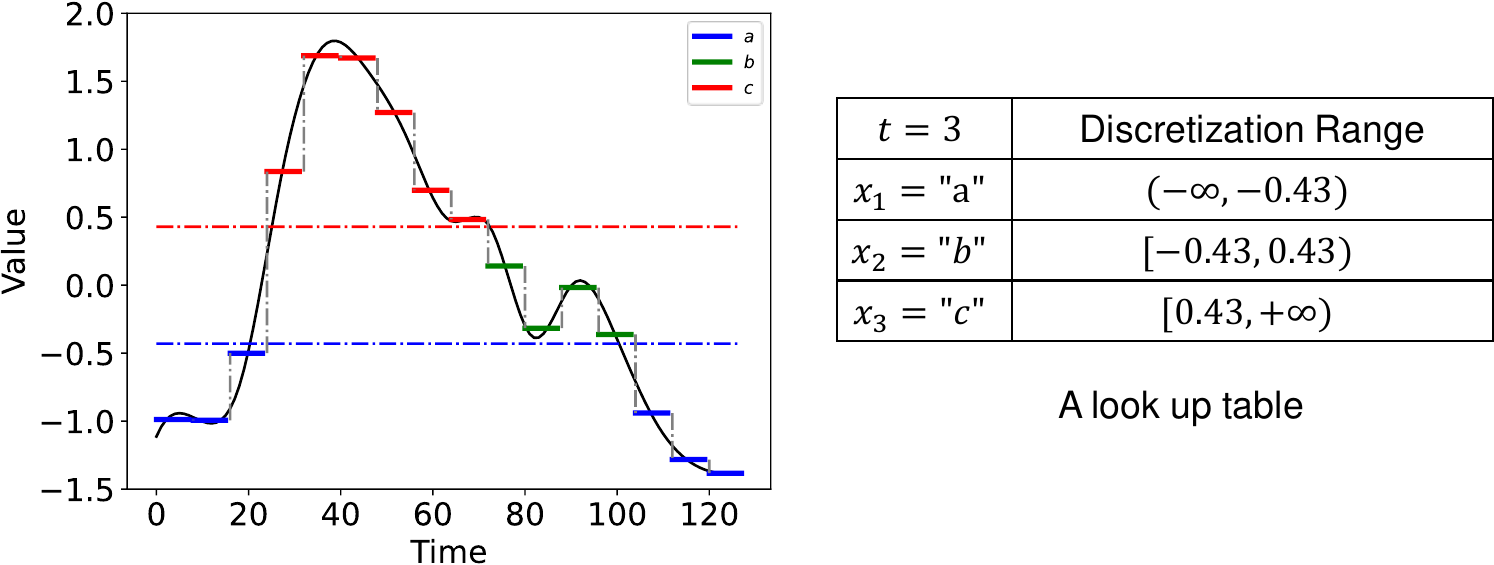}
	\vspace{-0.5em}
	\caption{A time series of length $m=128$ is compressed into a sequence ``aaaccccccbbbbaaa" with the segment length $w=8$ and symbol size $t=3$.}
	\label{sax_process}
	\vspace{-1.5em}
\end{figure}

An $m$-length time series $R$ is a sequence of $m$ values aligned with generated timestamps, denoted by
$R=\{r_1, r_2, \dots\, r_m\}$. 
To accommodate the setting of user-level privacy, it is necessary to reduce the dimensionality of the time series to ensure utility. Therefore, a pre-processing transformation~\cite{Lin2007, Luo2015, Chen2023} is required to transform a time series into shorter segments. In this paper, we utilize 
Symbolic Aggregate approXimation (SAX)~\cite{Lin2007} due to its compactness, effectiveness, and widespread usage~\cite{Li2020, Liu2022}. An example of SAX transformation is illustrated in Fig.~\ref{sax_process}. Given an $m$-length time series after z-score normalization, it is first segmented into $\lceil m/w\rceil$ pieces where $w$ is the segment length. Then the values in each segment are averaged, and all the averages form a vector $\bar{R}=\{\bar{r}_1, \bar{r}_2, \cdots, \bar{r}_{\lceil m/w\rceil}\}$, where $\bar{r}_i=\frac{1}{w}\sum_{j=w(i-1)+1}^{wi}r_j$.
Subsequently, given a symbol size $t$, each averaged value is assigned a symbol based on a lookup table~\cite{Lin2007}, see for example shown in Fig.~\ref{sax_process}. After SAX process, a time series is transformed into a shorter sequence $S=\{s_1, s_2, \dots\, s_{\lceil m/w\rceil}\}$, where each element $s_i$ is a symbol, and $s_i \in \{x_1, x_2, ..., x_t\}$. In the example, the utilized symbols are set as $x_1$=``a", $x_2$=``b", and $x_3$=``c".

\subsection{Local Differential Privacy}
Differential privacy (DP)~\cite{Dwork2006} was introduced to provide a privacy guarantee where there is a trusted data curator. 
To eliminate the assumption of a trusted data curator, local differential privacy (LDP)~\cite{Kasiviswanathan2008, Cormode2018, Ye2020} is proposed for data collection. With LDP, a user's data is locally perturbed before being uploaded to the data curator. Formally, LDP can be defined as follows. 

\begin{definition}[Local Differential Privacy] \label{def:ldp}
	Given a randomized mechanism $\mathcal{A}$, for any two inputs $v$ and $v'$, and any possible output $v^*$, the mechanism $\mathcal{A}$ satisfies $\epsilon$-local differential privacy ($\epsilon$-LDP) if and only if 
	\begin{equation}\nonumber
		\mathrm{Pr}(\mathcal{A}(v)=v^*)\le e^{\epsilon} \times \mathrm{Pr}(\mathcal{A}(v')=v^*).
	\end{equation} 
\end{definition}


In Def.~\ref{def:ldp}, the privacy budget $\epsilon$ controls the level of privacy, i.e., a smaller $\epsilon$ means a stronger privacy guarantee. When applying Def.~\ref{def:ldp} to time series data, $v$ and $v'$ refer to any two neighboring time series. In particular, the user-level neighboring time series is defined as follows. 

\begin{definition}[User-Level Neighboring Time Series] Given two time series $R$ and  $R'$, they are called user-level neighboring time series if and only if  the elements in $R$ and $R'$ are all different.
\end{definition}

The definitions of event-level privacy~\cite{Dwork2010} and $\omega$-event-level privacy~\cite{Kellaris2014} are similar to the user-level privacy. Specifically, any two event-level neighboring time series only differ in one element, while any two $\omega$-event level neighboring time series differ in $\omega$ consecutive elements. Both of them offer weaker privacy guarantees than user-level privacy.

\subsection{Problem Definition}
There are $n$ users, and each user holds a time series $R_i$. All these $n$ time series form a dataset $T=\{R_1, R_2, \cdots, R_n\}$, which is then transformed into a set of sequences $\hat{T}=\{S_1, S_2, \cdots, S_n\}$ by SAX. We aim to design an LDP mechanism $\mathcal{A}$ under user-level privacy to extract the top-$k$ frequent shapes in $\hat{T}$. In what follows, we first define frequent shape in Def.~\ref{def:shape}, based on which the top-$k$ frequent shapes are defined in Def.~\ref{def:top-k-shape}.

\begin{definition}[Frequent Shape]\label{frequentshape}
	\label{def:shape}
	Given a distance threshold $\theta$, a frequency threshold $N$, and a transformed dataset $\hat{T}$, a sequence $\mathcal{F}_i$ is a frequent shape if and only if its frequency among the dataset exceeds the threshold $N$, i.e., 
	\begin{equation}\nonumber
		|\{S_i|dist(\mathcal{F}_i, S_i)\le \theta, S_i\in \hat{T}\}|\ge N,
	\end{equation}
	where $dist(\cdot)$ is a distance measure between two sequences, such as the dynamic time warping (DTW) distance~\cite{Kate2016}.
\end{definition}

\begin{definition}[Top-$k$ Frequent Shapes]
	\label{def:top-k-shape}
	Given a set of frequent shapes $C =\{\mathcal{F}_1, \mathcal{F}_2, \cdots, \mathcal{F}_l\}$ ($l\ge k$) and a transformed dataset $\hat{T}$, let each transformed sequence $S_j\in \hat{T}$ find its closest match to a sequence $\mathcal{F}_r \in C$ by minimizing their distance, i.e., $\min_{\mathcal{F}_r\in C} dist(S_j, \mathcal{F}_r)$. A set of top-$k$ frequent shapes $\hat{C}=\{\hat{\mathcal{F}}_1,\hat{\mathcal{F}}_2, \cdots, \hat{\mathcal{F}}_k\}\subset C$, consists of $k$ sequences with the highest number of matches.
\end{definition}

Table~\ref{notation} summarizes the notations used in the paper.
\vspace{-1em}
\begin{table}[!htb]
	\vspace{-0.5em}
	\caption{List of notations.}
	\label{notation}
	\centering
	\renewcommand{\arraystretch}{1.1}
	\scriptsize{
		\begin{tabular}{c l}
			\hline
			\bfseries Notation &\bfseries Description \\ \hline
			$\epsilon$ & Privacy budget\\
			$t$ & Symbol size in SAX process\\
			$w$ & Segment length in SAX process \\
			$T$ & The time series dataset\\
			$\hat{T}$ & The sequences after SAX from $T$\\
			$S_i$ & The $i$-th sequence in $\hat{T}$\\
			$C$ & The estimated candidate shapes from all  users\\
			$\mathcal{F}_i$ & The $i$-th shape in C\\
			$u_i$ & The $i$-th user \\
			
			$\mathcal{F}_c$ & The  shapes sent by the server at each interaction\\
			\hline
	\end{tabular}}
	\vspace{-2em}
\end{table}

\section{Baseline Mechanism}\label{baseline}
This section first analyzes the challenges of designing a shape extraction mechanism that satisfies LDP under a user-level privacy guarantee. Then, we propose a baseline mechanism.

\vspace{-0.5em}
\subsection{Motivation}

Extracting valuable shapes from a noisy time series becomes particularly challenging under user-level LDP. To summarize, the difficulties are from two aspects. 

First, privacy budget allocation among all elements of a time series is infeasible as it may contain lots of elements or is even infinite. To ensure utility, previous mechanisms aim to reduce the elements that share a given privacy budget, such as the sampling-based mechanisms~\cite{Wang2020, Ahuja2023}. However, the utility is unsatisfactory when it comes to user-level privacy, since enormous sampled elements are required to retain the whole shape.
Furthermore, directly adding noise to the selected elements can significantly distort the shape.

Second, due to inherent LDP noise and data fluctuations in time series, it is crucial to filter out shape changes caused solely by them. The existing mechanism, PatternLDP~\cite{Wang2020}, encounters difficulties in selecting ``good" samples that can capture the trend changes from those noisy or fluctuating data.

Our baseline mechanism is proposed to address these challenges. To reduce processed elements, we adopt the SAX process and remove redundant information to compress time series, and then use a trie data structure to generate candidate shapes. To protect users' privacy, we perturb a user's selection from the generated shape candidates rather than directly adding noise to users' data. 

\subsection{Compressive SAX}
\begin{figure}[!htb]
	\centering
	\includegraphics[width=0.35\textwidth]{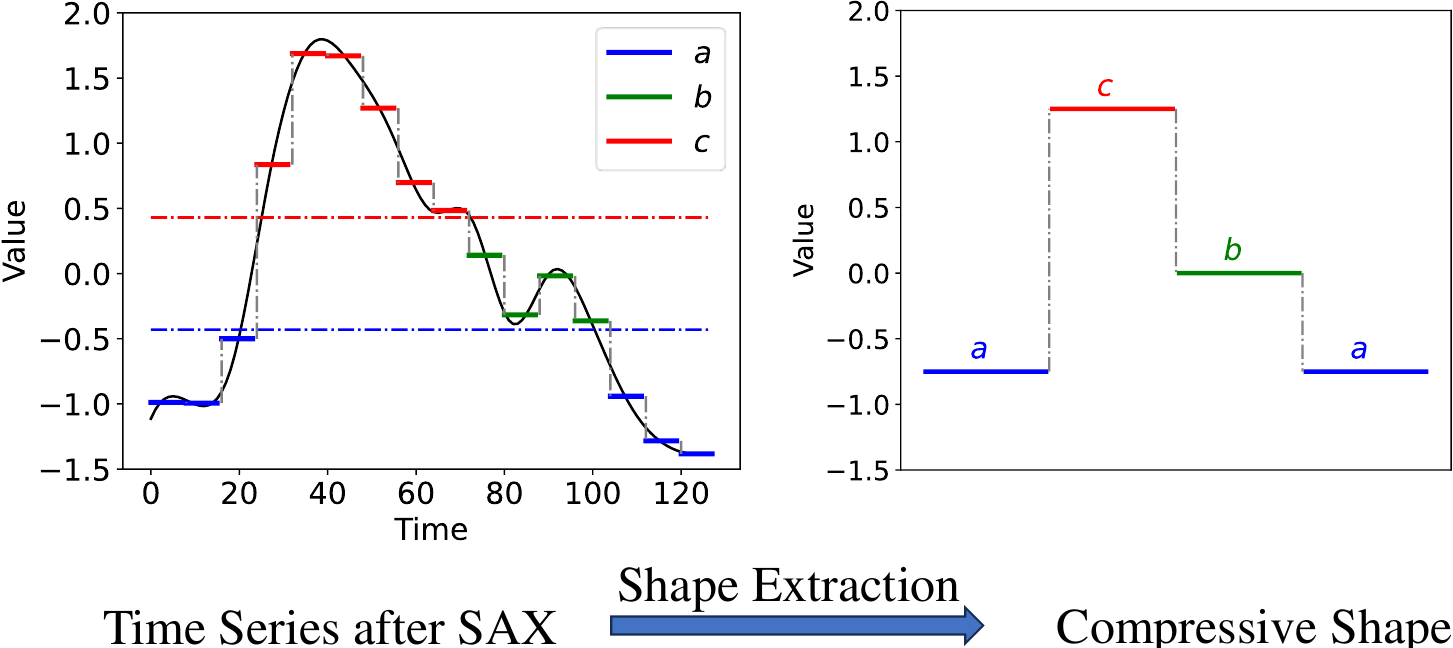}
	\caption{Extracted shape after Compressive SAX.}
	\label{essentialShape}
	
\end{figure}

To reduce the number of elements in a time series, we propose a dimension reduction method called {\it Compressive SAX}  by replacing simple sampling. 
As aforementioned, SAX transforms a time series into a shorter sequence while retaining much original shape information~\cite{Lin2007}. However, the sequence generated by SAX contains many repetitive symbols due to small data fluctuations. Since these repetitive symbols are not  essential to describe the shape, they can be merged as one. For instance, the trajectories of a gesture performed at different speeds share the same shape, so we can  merge those repetitive symbols in the sequences caused by slower speeds. 
Consequently, the sequence ``aaaccccccbbbbaaa" in Fig.~\ref{sax_process} can be reduced to ``acba", effectively reducing the number of elements while preserving the shape, as illustrated in Fig.~\ref{essentialShape}.

\vspace{-1.4em}
\subsection{Perturbation Mechanism}
Intuitively, we can directly perturb a time series' sequence generated by Compressive SAX. However, this may significantly distort its shape. Thus, we alternatively inject noise into the process of selecting shape candidates. 

To obtain the top-$k$ frequent shapes after Compressive SAX, we need to retain the elements, their frequencies, and their order in the sequences. A trie-based data structure fulfills all these requirements efficiently.
Unfortunately, the existing trie-based frequent sequence mining mechanisms such as PrivTrie\cite{Wang2018} and PEM\cite{Wang2021} are not suitable in our scenario.
Although PrivTrie is designed for frequent sequence mining, its practical deployment is challenging due to the substantial communication resources it requires~\cite{Wang2021}. On the other hand, PEM is specifically designed for frequent prefix mining of bits (i.e., 0 and 1), and it expands multiple levels in a single round to decrease the iterations, resulting in allocating more users in each iteration. However, in our scenario where the symbol size greatly exceeds two, directly applying PEM results in a large expansion domain, which greatly degrades utility. 

To this end, we have developed our baseline solution, which aims to optimize both efficiency and utility. 
Our solution consists of two steps: sequence length estimation and shape candidate generation. Initially, we randomly divide $n$ users into two groups: $P_a$ for length estimation and $P_b$ for mining frequent shape candidates. Sequence length estimation is to estimate the most frequent length $\ell_S$ of users' sequences, so as to determine the height of the trie. Shape candidate generation collects frequent sequences from users to facilitate the trie expansion. In each round of the trie expansion, $|P_b|/\ell_S$ users participate in this process, and each user contributes only once throughout the whole mechanism. 

\paragraph{Sequence Length Estimation}
Since the sequences are shortened by Compressive SAX, we can employ a frequency estimation mechanism to estimate the most frequent length $\ell_S$~\cite{Wang}.
Without loss of generality, we assume $\ell_S$  falls in a specific range, denoted by $[\ell_{low}, \ell_{high}]$.
For a sequence $S_{u_i}$ of a user $u_i$ from $P_a$, its length is initially truncated to fall into the range and then perturbed. The perturbation mechanism $\Phi(\cdot)$ can be any frequency estimation mechanism, such as Generalized Randomized Response (GRR) ~\cite{Wang2017}.
On the server side, the most frequent length $\ell_S$ can be aggregated by
\begin{equation}\label{lengthestimation}
	\ell_S = \argmax_{\ell_{low} \le \hat{\ell} \le  \ell_{high}, \hat{\ell}\in \mathbb{Z}}|\{u_i\mid \Phi(\ell_{u_i})=\hat{\ell}, u_i\in P_a\}|.
\end{equation}

\paragraph{Generation of Shape Candidates} In this step, we expand a trie to generate the top-$k$ frequent shapes. The trie starts with an empty node known as the ``root" and progressively grows to Level 1 based on the symbol size $t$. For candidate generation at each level, the population $P_b$ is randomly divided into $\ell_S$ groups, with each user group participating in one level. In what follows, we elaborate on this process from both the server side and the user side.

\begin{itemize}
	\item [1)]\textit{Trie expansion on the server side.} 
	Although it seems plausible to generate all the candidate  shapes with length $\ell_S$ at once, the following computation load for frequency estimation is extremely high because each user needs to compare their own sequence with thousands of candidates. Therefore, we propose an intuitive pruning step with a threshold at each level of trie expansion, and an example of trie expansion is shown in Fig.~\ref{trie}. In particular, the server first prunes the shape candidates estimated from a group of users at the current level with a threshold, i.e., Level 1. Afterward, Level 1 is expanded to the next level based on the symbol size $t$, with each leaf node establishing connections to $t-1$ child nodes. And then, the server sends the expanded candidates, starting from the root node to the corresponding group of users for frequency estimation. 
	\begin{figure}[!htb]
		\vspace{-1em}
		\centering
		\includegraphics[width=0.4\textwidth]{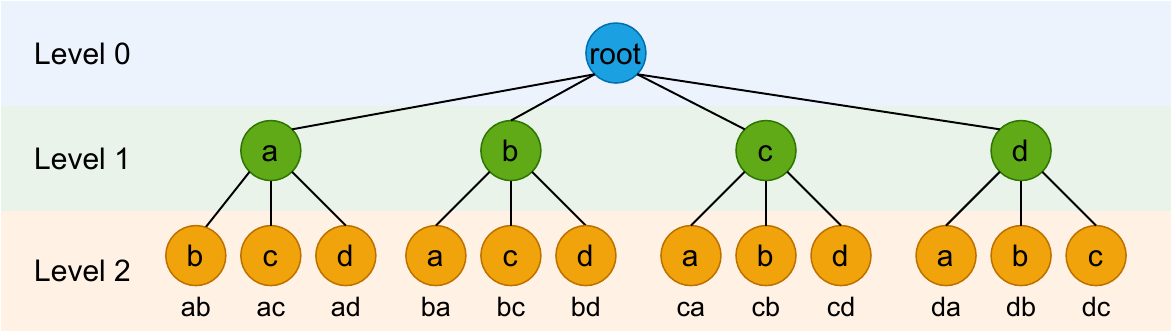}
		\caption{An example of trie expansion with symbol size $t=4$, and the symbols are ``a", ``b", ``c", and ``d". Moreover, if the estimated frequency of ``a", ``b", ``c", and ``d" exceeds the given threshold at Level 1, there is no need to prune the candidates  before its expansion to Level 2.}
		\label{trie}
		\vspace{-1em}
	\end{figure}

	\item [2)]\textit{Sequence matching on the user side.} After receiving the shape candidates from the server, each user follows a privacy-preserving selection process using the Exponential Mechanism (EM)~\cite{McSherry2007}. That is, each user chooses the most similar shape among the candidates, and then perturbs the selected candidate by EM. EM protects each user's privacy by releasing a shape similar to their own sequence with a higher probability and a different one with a lower probability.
	Therefore, the score function $\mathbb{S}(\cdot)$ in EM is related to a distance measure $dist(\cdot)$ between two sequences, i.e., $\mathbb{S}(\cdot)\propto \frac{1}{dist(\cdot)}$.
	Consequently, the score function $\mathbb{S}(\cdot)$  yields a large value when the two sequences are  similar, i.e., when the value of $dist(\cdot)$ is small.
	Given a user's sequence $ \mathcal{F}_{u_i}$ and $r$ shape candidates $\mathcal{F}_c = \{\mathcal{F}_{c_1}, \mathcal{F}_{c_2}, \cdots, \mathcal{F}_{c_r}\}$ sent by the server, the probability for the user to output a candidate $\mathcal{F}_{c_j}$ is
	\begin{equation}\label{emp}
		\mathrm{Pr}[\Psi(\mathcal{F}_{u_i})=\mathcal{F}_{c_j}]=\frac{\exp(\frac{\epsilon}{2\Delta}\mathbb{S}(\mathcal{F}_{u_i}, \mathcal{F}_{c_j}) )}{\sum_{\mathcal{F}_{c_z}\in \mathcal{F}_c}\exp(\frac{\epsilon}{2\Delta}\mathbb{S}(\mathcal{F}_{u_i}, \mathcal{F}_{c_z}))},
	\end{equation}
	where $\mathbb{S}(\cdot)$ is normalized to $[0, 1]$, and thereby EM leverages the sensitivity $\Delta=1$.
\end{itemize}
In the end, the server outputs the top-$k$ frequent shapes according to the estimated results at the leaf nodes. 


The baseline mechanism is summarized in Algorithm~\ref{match}. Initially, the users' time series are transformed by Compressive SAX and randomly split into two populations, denoted by $P_a$ and $P_b$. Subsequently, the trie height is estimated through a frequency estimation mechanism under LDP from the population $P_a$ (Lines 1-5). The trie is then expanded to generate the frequent shape candidates level by level from the population $P_b$ (Lines 6-12). During each level's expansion, the candidates are pruned before being distributed to the corresponding group of users (Line 7). The users allocated to a specific trie level select the most similar candidate, inject noise to the selection, and then upload the perturbed data (Lines 8-11). Ultimately, the server outputs the top-$k$ frequent shapes.

\begin{figure}
	\removelatexerror
	\begin{algorithm}[H]
		\caption{The baseline mechanism}
		\label{match}
		\begin{algorithmic}[1]
			\small
			\REQUIRE The privacy budget $\epsilon$, the sequences after Compressive SAX $\{S_1, S_2, \cdots, S_n\}$ from $n$ users, the length range of the sequences $\ell_{low}$ and $\ell_{high}$
			\ENSURE Top-$k$ frequent shapes $\{\hat{\mathcal{F}}_1, \hat{\mathcal{F}}_2, \cdots, \hat{\mathcal{F}}_k\}$
			
			\FOR{$u_i\in P_a$}
			\STATE Pad or truncate her sequence length $\ell_{u_i}$ into $[\ell_{low}, \ell_{high}]$
			\STATE Add noise $\hat{\ell}=\Phi(\ell_{u_i})$ by using a frequency estimation mechanism and send $\hat{\ell}$ to the server
			\ENDFOR
			\STATE The server derives the frequent length $\ell_S$ by Eq.~(\ref{lengthestimation})
			\FOR{level $\ell$ in $\{0, 1, \cdots, \ell_S-1\}$}
			\STATE The server first prunes the candidates at the current level and then expands to $(\ell+1)$-th level
			\FOR{$u_i\in P_b^{\ell_r}$}
			\STATE Calculate the distances between $S_i$ and  all candidates $\mathcal{F}_{c_j}\in \mathcal{F}_{c}$ sent by the server
			\STATE Select a candidate according to Eq.~(\ref{emp}) and send it back to the server.
			\ENDFOR
			\ENDFOR
			\STATE The server outputs the top-$k$ frequent shapes according to the estimated frequency at the leaf nodes.
		\end{algorithmic}
	\end{algorithm}
	\vspace{-2em}
\end{figure}

\subsection{Privacy Analysis}
\begin{theorem}\label{theo}
	The baseline mechanism satisfies $\epsilon$-LDP under user-level privacy.
\end{theorem}
\begin{IEEEproof}\label{proof}
	Note that Compressive SAX is a determinate process without randomness. For a time series $R$ and any user-level neighboring time series $R'$, the corresponding transformed sequences $S$ and $S'$ after Compressive SAX may be the same or different. For example, for any two identical input time series or the two time series bounded by the transformation difference, the transformed sequences will be the same. Otherwise, the transformed sequences will be different. And there will leave less privacy information after transformation compared with the original time series.
	
	If the corresponding transformed sequences $S$ and $S'$ after Compressive SAX are the same, we can easily prove the baseline mechanism to satisfy the privacy guarantee where $\epsilon=0$:
	\begin{equation}\nonumber
		\frac{\mathrm{Pr}[\Phi(S)=S^*]}{\mathrm{Pr}[\Phi(S')=S^*]}=\frac{\mathrm{Pr}[\Phi(S)=S^*]}{\mathrm{Pr}[\Phi(S)=S^*]}=\exp(0)\le \exp(\epsilon).
	\end{equation}
	In what follows, we consider the case when $S$ and $S'$ are different.
	
	As for length estimation,  an LDP mechanism is applied. Given $S$ and $S'$, and an LDP frequent estimation mechanism $\Phi(\cdot)$, there is
	\begin{equation}\nonumber
		\mathrm{Pr}[\Phi(|S|)=\ell^*]\le e^\epsilon\cdot\mathrm{Pr}[\Phi(|S'|)=\ell^*],
	\end{equation}
	where $\ell^*\in [\ell_{low}, \ell_{high}]$ is a possible output representing the sequence length.
	
	Moreover, the candidate generation also satisfies $\epsilon$-LDP under user-level privacy. Given  $S$, $S'$, and the Exponential Mechanism we utilized, there is
	\begin{equation}\nonumber
		\begin{split}
			&\frac{\mathrm{Pr}[\Psi(S)=\mathcal{F}_{c_i}]}{\mathrm{Pr}[\Psi(S')=\mathcal{F}_{c_i}]}=
			\frac{\frac{\exp(\frac{\epsilon}{2\Delta}\mathbb{S}(S, \mathcal{F}_{c_i}) )}{\sum_{\mathcal{F}_{c_z}\in \mathcal{F}_c}\exp(\frac{\epsilon}{2\Delta}\mathbb{S}(S, \mathcal{F}_{c_z}))}}{\frac{\exp(\frac{\epsilon}{2\Delta}\mathbb{S}(S', \mathcal{F}_{c_i}) )}{\sum_{\mathcal{F}_{c_z}\in \mathcal{F}_c}\exp(\frac{\epsilon}{2\Delta}\mathbb{S}(S', \mathcal{F}_{c_z}))}}\le\exp(\epsilon).
		\end{split}
	\end{equation}
	
	Therefore, the baseline mechanism satisfies $\epsilon$-LDP under user-level privacy by using the parallel composition theorem~\cite{McSherry2009}. 
\end{IEEEproof}


%
%


\section{PrivShape: An Optimized Mechanism}\label{improvement}
The baseline mechanism still faces a utility issue --- although we reduce the estimated candidates through a threshold-based pruning method, there are numerous expanded nodes in a single round of trie expansion, and the expansion domain grows exponentially, resulting in large perturbation domain that causes EM to experience utility loss. Moreover, selecting an appropriate threshold for pruning is challenging in practice --- a small threshold causes low pruning efficiency, whereas a large threshold may inadvertently prune some frequent shapes. In this section, we present an optimized mechanism {\it PrivShape} with two pruning strategies to enhance the baseline mechanism. The first strategy focuses on the expansion process for pruning, while the second strategy refines the top-$k$ frequent candidate set.

\subsection{Motivation}

Recall that in the Exponential Mechanism, if the perturbation domain is large, the probability that an element remains as itself after perturbation is small, thereby resulting in utility degradation. In the baseline mechanism, all node expansions are considered. For example, in Fig.~\ref{trie}, node ``a" at Level 1 expands to three \textbf{sub-shapes} (``a", ``b"), (``a", ``c"), (``a", ``d"), and then Level 1  expands to 12 nodes at Level 2.  As the trie grows, an enormous number of candidates will be expanded. Therefore, in addition to pruning the candidates at each level, we also need to reduce the expanded nodes to form the next level. As we stated before, a simple threshold-based pruning strategy is not feasible, so we propose a novel pruning strategy for trie expansion. Additionally, we also find that the results at the leaf nodes play a crucial role since the top-$k$ frequent shapes are obtained from these nodes. Hence, refining these results also becomes essential to enhance the utility of shape extraction.



\begin{figure}[!htb]
	\centering
	\includegraphics[width=0.48\textwidth]{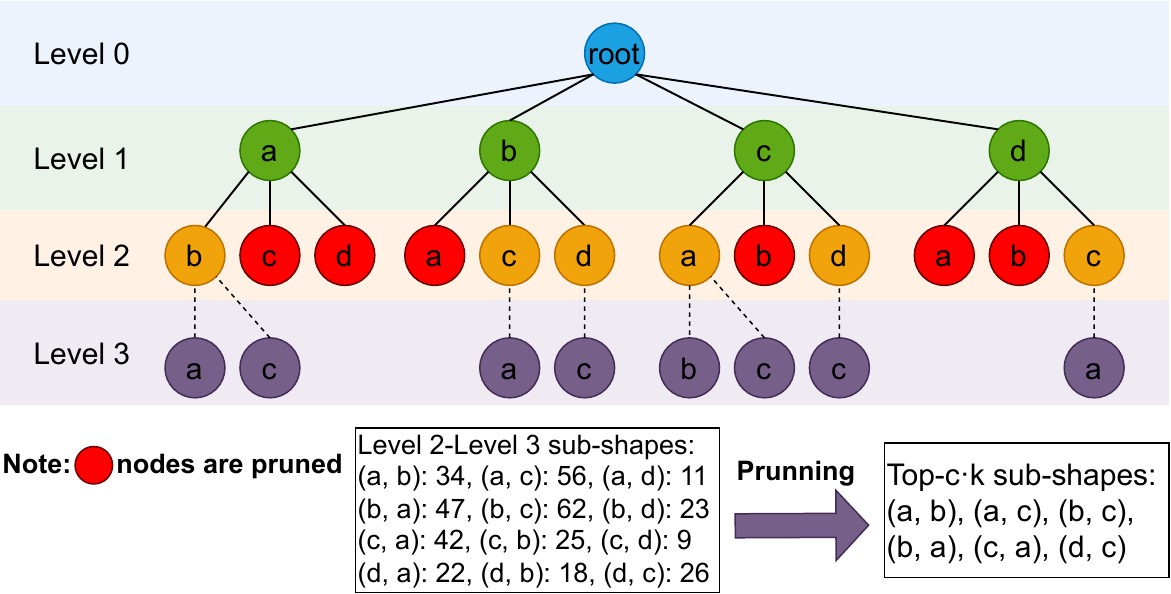}
	\caption{The trie expansion in PrivShape from Level 2 to Level 3 with $c=3$ and $k=2$. The candidates in Level 2 will be first pruned before expansion to Level 3. Moreover, the expansion candidates (i.e., Level 2-Level 3 sub-shapes) are also pruned.}
	\label{trie-expansion}
	\vspace{-1.5em}
\end{figure}

\subsection{Trie Expansion Strategy for Pruning}
In this subsection, we present a novel pruning strategy without a predefined threshold and the details are presented as follows. Drawing inspiration from Frequent Pattern Growth Algorithm~\cite{Han2000} that  sub-patterns of a frequent pattern are frequent, we only need to expand the frequent sub-shapes during each expansion. However, a sub-pattern is precisely matched with the frequent pattern in~\cite{Han2000}, but the matched shapes are measured by a distance metric in our scenario. For example, when given a distance threshold $\theta$, a sub-shape ``ab'' can be matched with another sub-shape ``ac''. Therefore, we need to prove that the sub-shape of a frequent shape is also frequent, as established in what follows.

Let a user's sequence $S$ be divided into two non-overlap subsequences, namely, the prefix $\text{PRE}_{S}$ and the suffix $\text{SUF}_{S}$. 
For ease of the proof, we assume that for any two sequences $S_i$ and $S_j$ with the same length,
$dist(S_i, S_j)=dist(\text{PRE}_{S_i}, \text{PRE}_{S_j})+dist(\text{SUF}_{S_i},\text{SUF}_{S_j})$,
where $|\text{PRE}_{S_i}|=|\text{PRE}_{S_j}|$.
Note that such an assumption is reasonable for distance measures, such as the Euclidean distance. Then in Lemma~\ref{lem1} we focus on prefixes, and the proof on suffixes can be derived in a same manner.  
\begin{lemma}\label{lem1}
	Given a distance threshold $\theta$, a frequency threshold $N$, and a dataset $\hat{T}$, a prefix $\text{PRE}_{\mathcal{F}_i}$ of a frequent shape $\mathcal{F}_i$ is also  frequent, i.e., 
	\begin{equation}\nonumber
		|\{S_i|dist(\text{PRE}_{\mathcal{F}_i}, \text{PRE}_{S_i})\le \theta, S_i\in \hat{T}\}|\ge N.
	\end{equation}
\end{lemma}
\begin{IEEEproof}
	According to Def.~\ref{frequentshape}, the frequency of a frequent shape $\mathcal{F}_i$ among the dataset $\hat{T}$ reaches the threshold $N$, denoted as
	\begin{equation}\nonumber
		|\{S_i|dist(\mathcal{F}_i, S_i)\le \theta, S_i\in \hat{T}\}|\ge N.
	\end{equation}
	In terms of a frequent shape $\mathcal{F}_i$ and its any matched sequence $S_i\in \hat{T}$, the prefix $\text{PRE}_{\mathcal{F}_i}$ is still matched with $S_i$ because the distance measurement is non-negative, i.e., $dist(\text{PRE}_{\mathcal{F}_i}, \text{PRE}_{S_i})\le dist(\mathcal{F}_i, S_i)\le \theta$.
	In summary, any prefix $\text{PRE}_{\mathcal{F}_i}$ of a frequent shape $\mathcal{F}_i$ is still frequent since
	\begin{equation}\nonumber
		\begin{split}
			&|\{S_i|dist(\text{PRE}_{\mathcal{F}_i}, \text{PRE}_{S_i})\le \theta, S_i\in \hat{T}\}|\\ &\quad\quad\quad\ge|\{S_i|dist(\mathcal{F}_i, S_i)\le \theta, S_i\in \hat{T}\}|\ge N.\\
		\end{split}
	\end{equation}
\end{IEEEproof}
\begin{theorem}
	Given a frequent shape $\mathcal{F}_i=\{s_1, s_2, \cdots, s_m\}$, its sub-shape $\text{SP}_j=\{s_j, s_{j+1}\},\quad j\in \{1, 2, \cdots, m-1\}$ is still frequent.
\end{theorem}
\begin{IEEEproof}
	Because a sub-shape $\text{SP}$ can be regarded as the suffix of a prefix extracted from the frequent shape $\mathcal{F}_i$, the sub-shape $\text{SP}$ is thus frequent according to Lemma~\ref{lem1}.
\end{IEEEproof}

In real-world scenarios, the assumption formulated in our proof should be relaxed as
\begin{equation}\nonumber
	dist(S_i, S_j)\le dist(\text{PRE}_{S_i}, \text{PRE}_{S_j})+dist(\text{SUF}_{S_i},\text{SUF}_{S_j}),
\end{equation}
where $|\text{PRE}_{S_i}|=|\text{PRE}_{S_j}|$, such as the dynamic time warping (DTW) distance, the string edit distance, and Hausdorff distance.  To avoid omitting some frequent shape candidates due to the assumption, we need to select more candidate sub-shapes instead of choosing only the top-$k$ candidates, where $k$ is the desired number of frequent shapes. In particular, for each round of trie expansion, we select top-$c\cdot k$ sub-shape candidates  where $c$ $(c\ge 2)$ is a constant.


To obtain the frequent sub-shapes, we estimate their frequency at each level from users by padding-and-sampling~\cite{Wang} after the length estimation. First, a user's sequence $S_i$ is padded or truncated to reach the length $\ell_S$, and then a sub-shape $(s_j, s_{j+1})$, $j \in \{1, 2, \cdots, \ell_S-1\}$, is randomly selected from the processed $S_i$. After padding-and-sampling, the user sends the chosen level $j$ and the noised sub-shape to the server, and the sent content is $(j, \Phi((s_j, s_{j+1})))$, where $\Phi(\cdot)$ represents an LDP perturbation mechanism (e.g., GRR). Finally, the server aggregates the sub-shapes at each level.


Additionally, we prune the candidates at leaf nodes for  next-level generation by top-$c\cdot k$ selection. Such a pruning process can enhance utility by reducing the candidates expanded to the next level, subsequently decreasing candidates at the following levels.
The trie-expansion process is illustrated in Fig.~\ref{trie-expansion}. First, at Level 2, we prune the candidates and only select the top-$c \cdot k$ (with $c=3$, $k=2$ in the example) candidates for expansion. Afterward, we choose the top-$c \cdot k$ frequent sub-shapes estimated from a group of users to expand  Level 2 to Level 3.

\subsection{Two-Level Refinement}\label{twolevel}
Since the leaf nodes play a crucial role in determining the final results, enhancing the estimated frequency of these nodes becomes necessary. In our mechanism, we allocate users for different tasks instead of dividing the privacy budget. Intuitively, we can  improve utility by allocating more users for a specific task, analogous to allocating more privacy budget~\cite{Wang2018}. 


To improve the estimated frequency at leaf nodes, we employ a two-level refinement~\cite{Wang2018a}. The first level is for frequent shape candidate estimation, and the other level is for candidate refinement. Besides randomly allocating the population for frequent length estimation with $P_a$ and sub-shape estimation with $P_b$, we randomly divide the remaining population into two parts, $P_c$ and $P_d$.  The first level is the  generation of candidate shapes by trie expansion from $P_c$, combining the trie expansion process in the baseline mechanism with the pruning strategies. While the second level is to prune the candidates at the leaf nodes through only selecting the top-$c\cdot k$ candidates  and re-estimate their frequency from $P_d$.

In addition, we note that there will output many similar shapes after processing the existing steps. When some similar candidate shapes are selected, the presence of the true candidates is concealed. To address this issue, we propose a post-processing strategy to avoid the selection of similar shapes. We first identify and group similar shapes together by partitioning the candidate shapes into $k$ clusters based on their distance measures.  Afterward, we select the most frequent candidate shape from each cluster to form the final result. This strategy ensures that only distinct shapes are chosen, promoting the discovery of actual top-$k$ frequent shapes.

PrivShape is summarized in Algorithm~\ref{privshape}. Compared with Algorithm~\ref{match}, the primary modification lies in the inclusion of the trie-expansion and the two-level refinement strategies. Following the transformation of users' time series using Compressive SAX, the processed sequences are randomly allocated to four groups, namely, $P_a$, $P_b$, $P_c$, and $P_d$.  The estimation (Line 1) of the most frequent sequence length from the users in $P_a$  is the same as the process in Algorithm~\ref{match} Lines 1-4. Subsequently, users from $P_b$ are employed for sub-shape estimation (Lines 2-6). Afterward, a trie is expanded to generate frequent shape candidates from $P_c$ (Lines 8-12). The generated candidates at leaf nodes are then pruned and re-estimated using users from $P_d$. Finally, the top-$k$ frequent shapes are output after the post-processing strategy to avoid selecting similar shapes (Line 15). 
\begin{figure}
	\removelatexerror
	\begin{algorithm}[H]
		\caption{PrivShape: An optimized mechanism}
		\label{privshape}
		\begin{algorithmic}[1]
			\small
			\REQUIRE The privacy budget $\epsilon$, the $n$ users' transformed sequences by Compressive SAX $\{S_1, S_2, \cdots, S_n\}$, the range of the sequence length $[\ell_{low}, \ell_{high}]$
			\ENSURE Top-$k$ frequent shapes $	\hat{C}=\{\hat{\mathcal{F}}_1,\hat{\mathcal{F}}_2, \cdots, \hat{\mathcal{F}}_k\}$
			
			\STATE Obtain the frequent sequence length $\ell_S$ from $P_a$
			\STATE $\#$ Frequent sub-shape estimation from $P_b$
			\FOR{user $u_i\in P_b$}
			\STATE The user $u_i$ randomly chooses a level $j\in \{1, 2, \cdots, \ell_S-1\}$ and reports a noised sub-shape with the level $(j, \Phi((s_j, s_{j+1})))$ by a randomized response LDP mechanism $\Phi(\cdot)$ 
			\ENDFOR
			\STATE The server aggregates the top-$c\cdot k$ sub-shapes at each level
			
			\STATE  $\#$ The trie expansion with $|P_c|/\ell_s$ users at each level
			\FOR{ each level $\ell\in \{0, 1, \cdots, \ell_S-1\}$}
			\STATE The server expands the trie to next level by the top-$c\cdot k$ sub-shapes at that level and then sends the candidate shapes to the corresponding group of users
			\STATE Each user allocated for the $\ell$-th level expansion selects and uploads a candidate shape based on Eq.~(\ref{emp})
			\STATE The server aggregates the uploaded selections and  prunes the candidates within top-$c\cdot k$ for next level expansion
			\ENDFOR
			\STATE $\#$ The two-level refinement from $P_d$
			\STATE Prune the leaf nodes and re-estimate the candidates
			\STATE Employ the post-processing strategy to output the top-$k$ frequent shapes $	\hat{C}=\{\hat{\mathcal{F}}_1,\hat{\mathcal{F}}_2, \cdots, \hat{\mathcal{F}}_k\}$
		\end{algorithmic}
	\end{algorithm}
	\vspace{-3em}
\end{figure}

\subsection{Privacy Analysis}
\begin{theorem}
	\label{thm:privshape}
	PrivShape satisfies $\epsilon$-LDP under user-level privacy.
\end{theorem}
\begin{IEEEproof}
	Compressive SAX is a determinate process devoid of randomness. Given a time series $R$ and any neighboring time series $R'$ at the user level, the  transformed sequences $S$ and $S'$ after Compressive SAX may either be identical or different. When considering two identical input time series or those whose transformation difference falls within a specific bound, the transformed sequences will be identical. Conversely, if the time series differ beyond this bound, the transformed sequences will differ as well. Additionally, the transformation process tends to reduce the  private information compared to the original time series.
	
	If the corresponding transformed sequences $S$ and $S'$ by Compressive SAX are the same, referring to Theorem 1, we can easily prove PrivShape satisfies $\epsilon$-LDP under user-level privacy where $\epsilon=0$. 
	
	Subsequently, we consider the case when $S$ and $S'$ are different. 
	First, we need to emphasize that the pruning strategy is conducted after processing an LDP mechanism. Thus, it does not violate privacy due to the post-processing theorem~\cite{Li2017}.
	Referring to Theorem 1, the length estimation, the shape candidates generation from trie expansion, and the two-level refinement can be easily proved to satisfy $\epsilon$-LDP under user-level privacy.
	
	As for sub-shape estimation, it satisfies the privacy guarantee because an LDP mechanism $\Phi(\cdot)$ is applied. Given a randomly chosen level $j\in \{1, 2, \cdots, \ell_S-1\}$, two sub-shapes $(s_j, s_{j+1})$ truncated at the $j$-th position from $S$ and $(s'_j, s'_{j+1})$ truncated from $S'$, there is
	\begin{equation}\nonumber
		\mathrm{Pr}[\Phi(s_j, s_{j\!+\!1})=(s^*_j, s^*_{j\!+\!1})]\le e^\epsilon \times \mathrm{Pr}[\Phi(s'_j, s'_{j\!+\!1})=(s^*_j, s^*_{j\!+\!1})],
	\end{equation}
	where $(s^*_j, s^*_{j+1})$ is a possible output from the $t(t-1)$ combinations of the $t$ symbols $\{x_1, x_2, \cdots, x_t\}$.
	
	Therefore,  PrivShape satisfies $\epsilon$-LDP under user-level privacy according to the parallel composition theorem~\cite{McSherry2009}. 
\end{IEEEproof}

\subsection{Utility Analysis}

\begin{theorem}
	In terms of the $\ell$-th ($\ell\ge 1$) level trie expansion, PrivShape improves the utility compared with the baseline mechanism by $\frac{t(t-1)^{\ell-1}}{c^2k^2}$ when given the symbol size $t$ in the worst case.
\end{theorem}
\begin{IEEEproof}
	According to the utility theorem~\cite{Dwork2014}, when given a reachable score $c$, the utility  for the Exponential Mechanism $\mathcal{M}_E$ can be denoted as
	\vspace{0em}  
	\begin{equation}\nonumber
		\mathrm{Pr}[u(\mathcal{M}_E(x, u, \mathcal{R})\le c]\le \frac{\left|\mathcal{R}\right|}{\left|\mathcal{R}_{\mathrm{OPT}}\right|}\exp(\frac{\epsilon(c-\mathrm{OPT}_u(x))}{2\Delta u}),
	\end{equation}
	\vspace{0em}
	where $x$ is the input instance for perturbation, $u$ is the score function,  $\mathcal{R}$ is the perturbation domain, and $\mathcal{R}_{\mathrm{OPT}}=\{r\in \mathcal{R}\mid u(x)=\mathrm{OPT}_u(x)\}$, $\left|\cdot\right|$ denotes the cardinality. 
	In our framework, the utility score is normalized, thus $\Delta u=1$. Moreover, $\mathrm{OPT}_u(x)$ is $\max u(x, r)=1, \forall r\in \mathcal{R}$. We have the probability
	\vspace{-0.2em}  
	\begin{equation}\nonumber
		\mathrm{Pr}[u(\mathcal{M}_E(x, u, \mathcal{R})\le c]
	\le |\mathcal{R}|\exp(\frac{\epsilon(c-1)}{2}),
\end{equation}
\vspace{-0.2em}  
with $|\mathcal{R}_{\mathrm{OPT}}|\ge 1$.

For the baseline mechanism $\mathcal{M}_E^B(x, u, \mathcal{R}^B)$, the utility is denoted as 
\begin{equation}\nonumber
	\mathrm{Pr}[u(\mathcal{M}_E^B(x, u, \mathcal{R}^B)\le c]\le |\mathcal{R}^B|\exp(\frac{\epsilon(c-1)}{2}),
\end{equation}
where $|\mathcal{R}^B|$ is the corresponding perturbation domain size.
Similarly, the utility  for PrivShape $u(\mathcal{M}_E^P(x, u, \mathcal{R}^P)$ can be denoted as 
\begin{equation}\nonumber
	\mathrm{Pr}[u(\mathcal{M}_E^P(x, u, \mathcal{R}^P)\le c]\le |\mathcal{R}^P|\exp(\frac{\epsilon(c-1)}{2}).
\end{equation}
Therefore, the utility improvement can be denoted as 
\begin{equation}\nonumber
	\frac{\mathrm{Pr}[u(\mathcal{M}_E^B(x, u, \mathcal{R}^B)\le c]}{\mathrm{Pr}[u(\mathcal{M}_E^P(x, u, \mathcal{R}^P)\le c]}\le \frac{|\mathcal{R}^B|}{|\mathcal{R}^P|}.  
\end{equation}
Since $\frac{|\mathcal{R}^B|}{|\mathcal{R}^P|}$ is obtained after the pruning process, it is highly correlated with the data. Here we can only analyze the worst-case scenario where neither PrivShape nor the baseline mechanism can be pruned effectively. We have 
$ \frac{|\mathcal{R}^B|}{|\mathcal{R}^P|}\le \frac{t(t-1)^{\ell-1}}{c^2k^2}.$  

\end{IEEEproof}

Additionally, the overall improvement can be denoted as
\begin{equation}\nonumber
\frac{\mathrm{Pr}[u(\mathcal{M}_E^B(x, u, \mathcal{R}^B)\le c]}{\mathrm{Pr}[u(\mathcal{M}_E^P(x, u, \mathcal{R}^P)\le c]}
\le \frac{\sum_{\ell_i=1}^{\ell_S}|\mathcal{R}^B_{\ell_i}|}{\sum_{\ell_i=1}^{\ell_S}|\mathcal{R}^P_{\ell_i}|}
\le\frac{t(t-1)^{\ell_S}-t}{\ell_Sc^2k^2(t-2)}.  
\end{equation}

Based on the utility analysis presented above, it is clear that reducing the perturbation domain size in the exponential mechanism holds significant importance. This outcome serves as theoretical proof of the enhancements achieved in PrivShape compared to the baseline mechanism. 

\subsection{Complexity Analysis}

We analyze the complexity of the baseline mechanism and PrivShape. Since the operation amount is highly correlated with the data, we can only consider the worst-case scenario where both PrivShape and the baseline mechanism cannot be pruned effectively.

\subsubsection{The baseline mechanism} In spite of length estimation process, the time complexity of a user is $O(1)$ for perturbation. For the server, the time complexity is $O(|P_a|)+O(\ell_{high}-\ell_{low})$ to collect perturbed lengths and find the frequent length. In the $\ell$-th trie expansion, the time complexity of a user is $O(\mathcal{M}(\ell)t(t-1)^{\ell-1})$ to find the candidate for uploading, where $\mathcal{M}(\ell)$ is the time complexity of distance measure between two $\ell$-length strings.
Moreover, the time complexity of the server is $O(|P_b|/\ell_{S})+O(t(t-1)^\ell)$ to collect the perturbed candidates and expand the nodes to next trie level. And it consumes $O(t(t-1)^{\ell_S-1}\log(t(t-1)^{\ell_S-1}))$ to find the top-$k$ candidates for output.
\subsubsection{PrivShape} The time complexity of a user in length estimation and sub-shape estimation is $O(1)$ for perturbation. The time complexity of the server in length estimation is $O(|P_a|)+O(\ell_{high}-\ell_{low})$, and is $O(|P_b|)+O(\ell_{S}t(t-1)\log(t(t-1)))$ in sub-shape estimation. For the $\ell$-th trie expansion, the time complexity of a user is $O(\mathcal{M}(\ell)c^2k^2)$ where $\mathcal{M}(\ell)$ is the time complexity of distance measure between two $\ell$-length strings. And the time complexity of the server is $O(|P_c|/\ell_S)+O(c^2k^2\log(ck))$. Finally, the time complexity in the two-level refinement is $O(\mathcal{M}(\ell_S)ck)$ for a user where $\mathcal{M}(\ell_S)$ is the time complexity of distance measure between two $\ell_S$-length strings, and is $O(|P_d|)+O(c^2k^2\log (ck))$ for the server.

In rough terms, the time complexity of PrivShape is expected to be better than that of the baseline mechanism, as the complexity of the baseline mechanism may involve exponential factors. The advantage of PrivShape stems from its effective pruning strategy. 




\section{Experimental Evaluation}\label{experiment}

\begin{table}[!htb]
	\vspace{-2em}
	\caption{The Details of the Datasets.}
	\label{dataset}
	\centering
	\renewcommand{\arraystretch}{1.5}
	\scriptsize{
		\begin{tabular}{|m{1.5cm}<{\centering}|m{2cm}<{\centering}|m{2cm}<{\centering}|m{1.5cm}<{\centering}|}
			\hline
			\bfseries Datasets &Symbols &  Trace &Trigonometric Wave \\ \hline
			\bfseries Task& Clustering & Classification & Classification \\ \hline
			\bfseries Length&398&275&Varying \\\hline
			\bfseries Original Instances & 1020&50&/ \\\hline
			\bfseries Generated Instances & 40,000& 40,000&20,000 \\\hline
			\bfseries Generation Method& \multicolumn{1}{m{2cm}|}{Generative Adversary Network with Bidirectional Long Short-Term Memory~\cite{Paszke2019}}& \multicolumn{1}{m{2cm}|}{The same generative model as the original work~\cite{Roverso2000}} & \multicolumn{1}{m{1.5cm}|}{Sampled points within one period}\\ \hline
			
	\end{tabular}}
	\vspace{-2em}
\end{table}

In this section, we demonstrate the superiority of our proposed mechanisms through experiments.
To  evaluate the effectiveness of our shape extraction mechanisms practically, we conduct experiments on two real applications, namely times series clustering and classification\textsuperscript{*}.\footnote{* Our code is available at https://github.com/Abigail-MAO/ICDE\_code}

\vspace{-0.5em}
\subsection{Dataset}
\begin{figure}[!htb]
	\centering  
	\subfigure[Symbol Dataset.]{
		\includegraphics[width=0.15\textwidth]{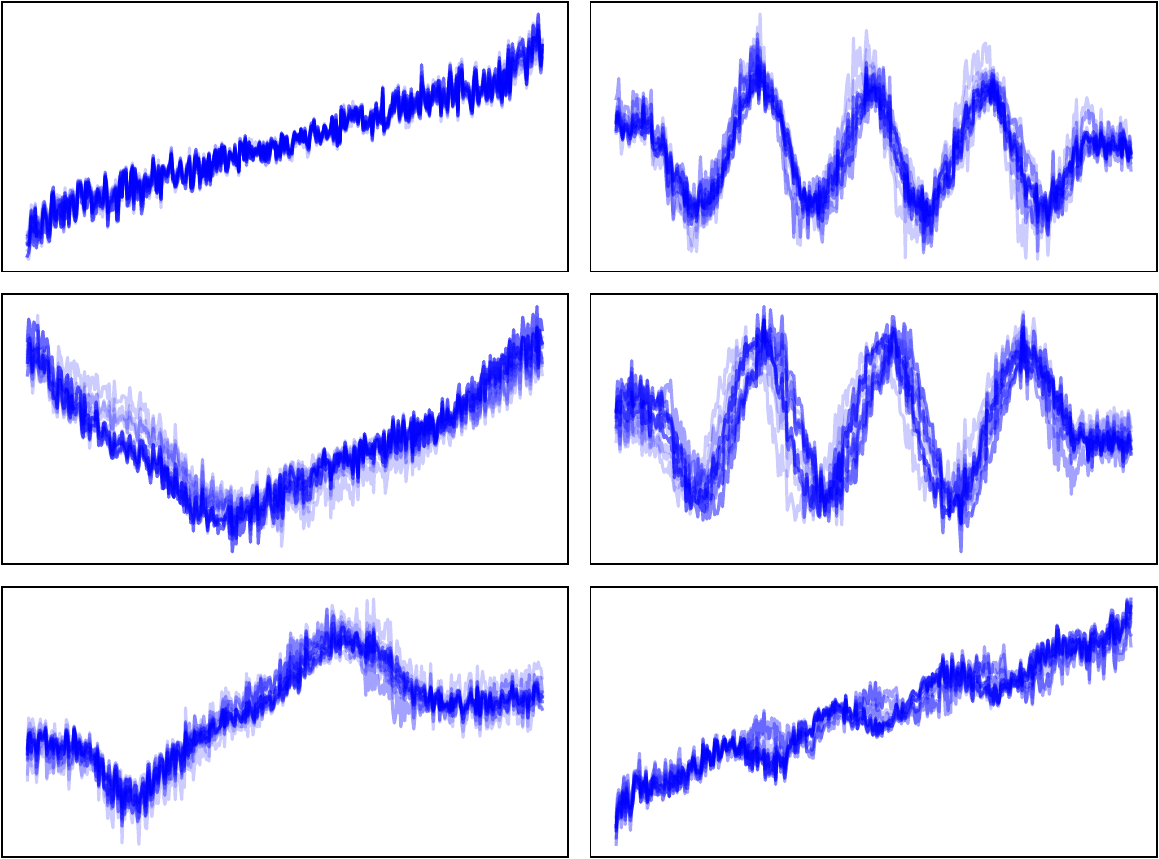}}
	\hspace{0.015\textwidth}
	\subfigure[Trace Dataset.]{
		\includegraphics[width=0.15\textwidth]{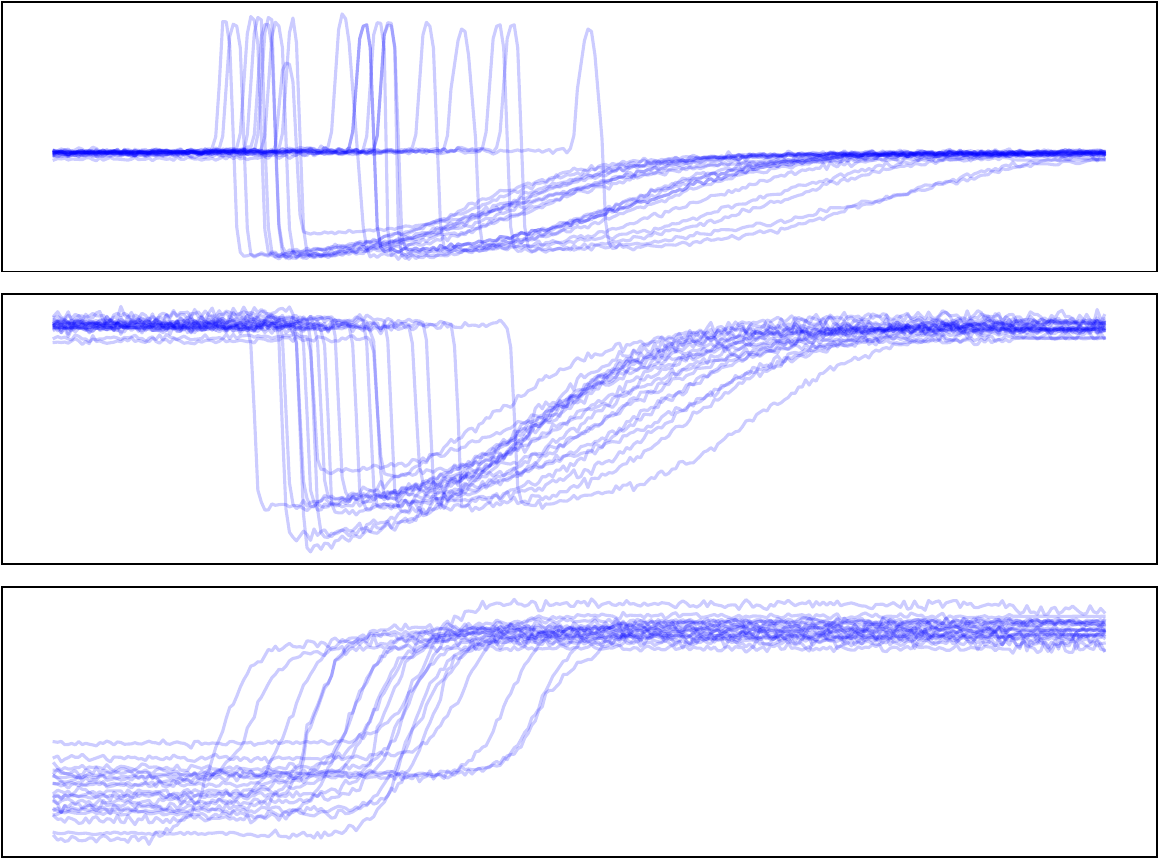}}
	\caption{Shapes from 20 randomly selected instances per class are depicted. }
	\label{shapeinfo}
	\vspace{-1em}
\end{figure}

In our experiments, we utilize two benchmark datasets, Symbols and Trace, from the UCR time series classification dataset~\cite{dataset}. Moreover, both datasets are already z-score normalized. The Symbols dataset records the x-axis motion of a user's hand in shape drawing experiments. These motion trajectories are categorized into six classes. As for the Trace dataset, it is obtained from the monitoring devices in a nuclear station, and we select three classes, which is the same as~\cite{Tavenard2020}. To further augment both two datasets with more instances, we adopt the generative models to generate 40,000 additional instances for each dataset. The shape information of Symbols and Trace datasets can be found in Fig.~\ref{shapeinfo}, both of which exhibit similar essential shapes within the same class. To assess the performance of the mechanisms under varying time series lengths, we generate Trigonometric Wave datasets, namely, sine and cosine values within one period for classification.  Further details of the datasets are provided in Table~\ref{dataset}. 

\subsection{Experiment Design} We perform two sets of experiments. The first set assesses the utility of shape extraction for clustering and classification. The second set explores the utility under various settings.
\subsubsection{Compared Method}
In our comparative analysis, we evaluate our proposed mechanisms against the value perturbation method. Namely, we compare with the existing shape-retaining mechanism PatternLDP~\cite{Wang2020} by extending it to satisfy user-level LDP and accommodate the privacy budget allocation strategy for offline use. We allocate the private budget for the sampled points according to their importance score, which is derived from the PID control error~\cite{Wang2020}. A larger PID control error indicates a significant change at that sampled point, and a larger privacy budget will be allocated.
As for the LDP mechanism for frequent length estimation and frequent sub-shape estimation, we use GRR~\cite{Wang2017} in the experiment.

\subsubsection{Distance Metrics}
We practically choose the distance metrics that yield reasonable performance on each dataset.  The distance metric used for the Symbols dataset is the dynamic time warping (DTW) distance~\cite{Kate2016}, while the Trace dataset is the string edit distance (SED)~\cite{Lin2007}.  Further investigations of distance metrics will be presented in the following experiments.

\subsubsection{Parameter Setting}
As for the parameters, we set $P_a=0.02n$, $P_b=0.08n$, $P_c=0.7n$, and $P_d=0.2n$, where $n$ is the number of time series in a dataset. In detail, we utilize 2\% users to estimate the frequent sequence length and 8\%  users for frequent sub-shape estimation. The trie expansion process consumes 70\% users and the final two-level refinement utilizes 20\% users.  The sequence lengths after Compressive SAX are then clipped by $\ell_{low}=1$,  $\ell_{high}=10$ for Trace dataset and $\ell_{high}=15$ for Symbols dataset. The constant $c$ to select the top-$c\cdot k$ candidates is set as $c=3$ for frequent sub-shape estimation and trie expansion. Moreover, in the baseline mechanism, we set a pruning threshold $N=100$, which means that the candidate sequences at each round of trie expansion with a frequency smaller than 100 are pruned. 

All the mechanisms are implemented in Python and are conducted on a server with Intel(R) Xeon(R) CPU E5-2620 v3 2.40GHz CPU, 20 cores, 128G RAM, running Ubuntu operating system. All the results are averaged over 500 trials.

\subsection{Metrics}
Since KMeans is  a state-of-the-art time series clustering algorithm~\cite{Javed2020}, we combine PatternLDP and  KMeans with the default settings~\cite{Pedregosa2011} to find the clusters. To assess the results, we employ the Adjusted Rand Index ($\mathrm{ARI}$)~\cite{Hubert1985, Pedregosa2011} as the evaluation metric. The $\mathrm{ARI}$ value ranges from $-1$ to $1$, with higher values indicating better clustering results. If $\mathrm{ARI}=0$, the clusters tend to be randomly generated. \textbf{For the Symbols dataset without LDP noise, the $\mathrm{\textbf{ARI}}$ of KMeans clustering is $1$, which can be regarded as the ground-truth.} For the baseline mechanism and PrivShape, we set the obtained top-$k$ frequent shapes as the cluster centroid.
And the distance metric we utilized in the clustering task is the dynamic time warping (DTW) distance~\cite{Kate2016}.

In the time series classification task, we evaluate the results based on the classification accuracy. We combine PatternLDP and the random forest classifier (RF) with the default settings~\cite{Pedregosa2011}, which is known for its fast performance and high accuracy. \textbf{Moreover, RF achieves a remarkable 100\% accuracy on the Trace dataset without LDP noise.} For the baseline mechanism and PrivShape, we utilize the most frequent shapes estimated within each class as the classification criteria. 
We employ the string edit distance (SED)~\cite{Lin2007} as the distance metric.

\begin{figure}[!htb]
	\vspace{0em}
	\centering
	\includegraphics[width=0.48\textwidth]{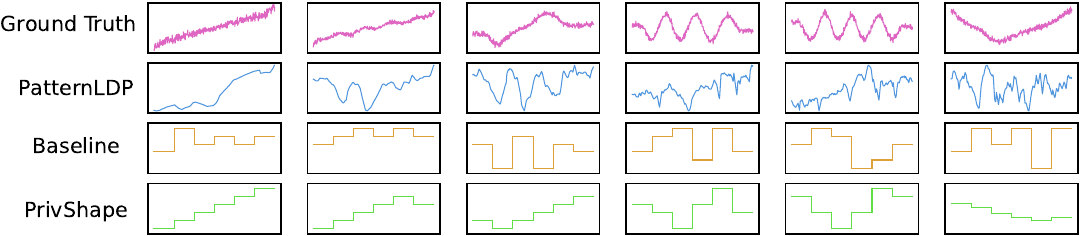}
	\caption{The shapes of Ground Truth and PatternLDP are obtained from the original dataset and the perturbed dataset generated by PatternLDP, respectively. After perturbation, we utilize KMeans~\cite{Pedregosa2011} to obtain the shape centers and match the centers by DTW distance~\cite{Rakthanmanon2012,Giorgino2009}. The shapes of Baseline and PrivShape are derived from the baseline mechanism and PrivShape with parameters  $w=25$ and $t=6$. And the corresponding matches with Ground Truth are determined using DTW distance~\cite{Kate2016}.}
	\label{motionShape}
	\vspace{-1.3em}
\end{figure}

\subsection{Clustering Task}

We plot the extracted shapes with $\epsilon=4$ in Fig.~\ref{motionShape} from once experiment, setting the random seed as 2023.  We adopt the same parameters of PatternLDP as in its original paper~\cite{Wang2020}. 
According to Fig.~\ref{motionShape}, the shapes of PatternLDP are nearly random generated. In contrast,  the shapes of PrivShape are more similar to Ground Truth. We also utilize quantitative measures to present the similarity among the extracted shapes. Since our derived shapes are strings, whereas the shapes of PatternLDP and Ground Truth are numerical, we first transform the shapes of Ground Truth and PatternLDP using Compressive SAX with the same settings as PrivShape. We then measure the distances between the derived shapes from  Ground Truth. The results shown in Table III  demonstrate the superior performance of our method.
\begin{table}
	\centering
	\caption{Quantitative measures of shapes (Symbols).}
	\renewcommand{\arraystretch}{1.5}
	\scriptsize{
		\begin{tabular}{|c|c|c|c|c|c|}
			\hline
			\textbf{Task}&\textbf{Mechanism}&\textbf{DTW} & \textbf{SED} & \textbf{Euclidean}&\textbf{ARI}\\
			\hline
			\multirow{3}*{Clustering}&PatternLDP & 38.97& 10.11&46.3&0.00 \\ \cline{2-6}
			&Baseline & 32.74& 12.81&35.86 &0.45\\  \cline{2-6}
			&PrivShape & 20.99& 1.83&4.74 &0.68\\ 
			\hline
	\end{tabular}}
	\label{quantiSymbols}
	
	\vspace{-2em}
\end{table}

\begin{figure}[!htb]
	\centering
	\includegraphics[width=0.3\textwidth]{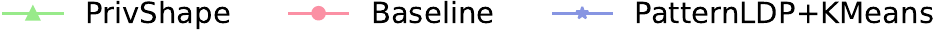}
	\includegraphics[width=0.28\textwidth]{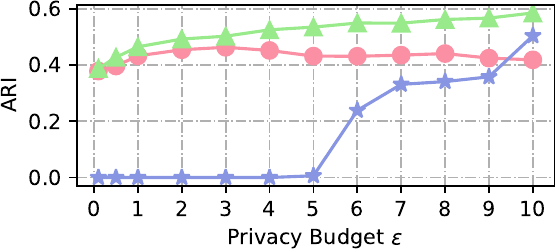}
	\caption{Clustering results on Symbols dataset varying $\epsilon$.}
	\label{motionResult}
	\vspace{-1em}
\end{figure}

In Fig.~\ref{motionResult}, we present the clustering results with varying privacy budgets $\epsilon=0.1, 0.5, 1, 2, \cdots, 10$.  According to Fig.~\ref{motionResult}, the utility of PatternLDP is insignificant even when $\epsilon=4$. This outcome is attributed to the noisy data and its selection of too many elements. As such, each selected element is allocated a small privacy budget, resulting in significant shape distortion. Conversely, our proposed mechanism PrivShape consistently outperforms PatternLDP.  

\vspace{-0.5em}

\subsection{Classification Task}
Because the classification task requires consideration of class labels, we modify the second level of the two-level refinement in Section~\ref{twolevel} to perturb a user's sequence with the corresponding label  using another LDP mechanism, the Optimized Unary Encoding (OUE) mechanism. In this case, the number of the encoding cells in OUE is $ck^2$, i.e., $ck$ candidates and $k$ classes. 

\begin{figure}[!htb]
	\vspace{-1em}
	\centering
	\includegraphics[width=0.27\textwidth]{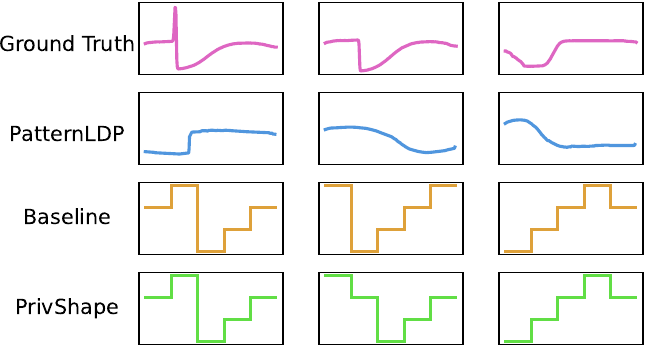}
	\caption{The shapes of Ground Truth and PatternLDP are obtained from the original dataset and the perturbed dataset generated by PatternLDP, respectively. We first utilize KShape~\cite{Paparrizos2015} (KShape is suitable to capture shapes from time series that are not warping) to get the shape centers. And then, we match the centers by DTW distance~\cite{Rakthanmanon2012,Giorgino2009}. And the shapes of Baseline and PrivShape are derived from the baseline mechanism and PrivShape with  $w=10$ and $t=4$. Note that the baseline mechanism and PrivShape output the shapes with corresponding labels.} 
	\label{traceShape}
	\vspace{-0.5em}
\end{figure}

In Fig.~\ref{traceShape}, we plot the extracted shapes from once experiment on the Trace dataset with $\epsilon=4$ while setting the symbol size $t=4$ and the segment length $w=10$ for Compressive SAX. The parameters of PatternLDP remain consistent with its original paper~\cite{Wang2020}. The random seed is set as 2023. Our optimized mechanism,  PrivShape, effectively captures the frequent shapes, whereas the shapes extracted by PatternLDP significantly differ  from Ground Truth. The quantitative similarity measures via the distances are shown in Table~IV. Note that Ground Truth and PatternLDP are also pre-processed by Compressive SAX in the same setting as PrivShape. The quantitative measures validate the shape similarity presented in Fig.~\ref{traceShape}.

\begin{table}
	\centering
	\caption{Quantitative measures of shapes (Trace).}
	\renewcommand{\arraystretch}{1.5}
	\scriptsize{
		\begin{tabular}{|c|c|c|c|c|c|}
			\hline
			\textbf{Task}&\textbf{Mechanism}&\textbf{DTW} & \textbf{SED} & \textbf{Euclidean}& \textbf{Accuracy}\\
			\hline
			\multirow{3}*{Classification}&PatternLDP & 17.42& 7.70&6.70&0.18 \\ \cline{2-6}
			&Baseline & 12.06& 3.34&5.90 & 0.85\\ \cline{2-6}
			&PrivShape & 12.06& 2.67&4.89& 0.87\\ 
			\hline
	\end{tabular}}
	\label{quantiTrace}
	
	\vspace{-2em}
\end{table}

\begin{figure}[!htb]
	\centering
	\includegraphics[width=0.3\textwidth]{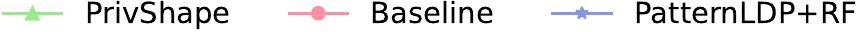}
	\vspace{0em}
	
	\includegraphics[width=0.3\textwidth]{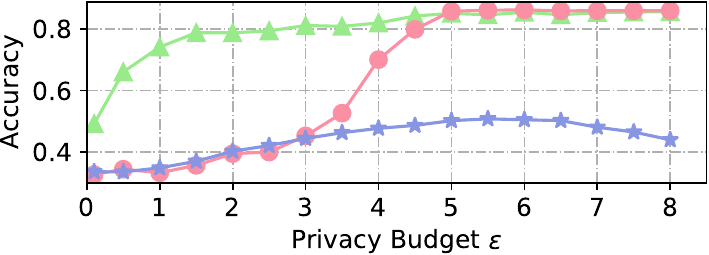}
	\caption{Classification results on Trace dataset.}
	\label{traceResult}
	\vspace{-1.5em}
\end{figure}

\begin{figure}[!htp]
	\centering
	\includegraphics[width=0.27\textwidth]{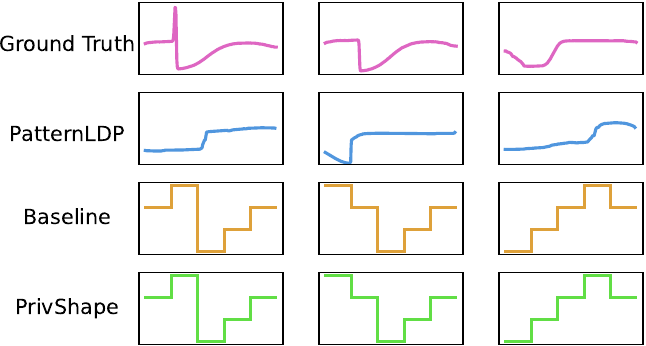}
	\caption{The extracted shapes from Trace dataset, and the settings are the same as Fig.~\ref{traceShape} except for $\epsilon=8$.}
	\label{pattern8}
	\vspace{-1em}
\end{figure}

In Fig.~\ref{traceResult}, we illustrate the classification results with varying privacy budgets $\epsilon=0.1, 0.5, 1, 1.5, \cdots, 8$.   Notably,  PrivShape consistently outperforms PatternLDP, even with small privacy budgets ($\epsilon\le2$). Additionally, we observe that the classification accuracy of PatternLDP first increases and then decreases when $\epsilon>6$. Such a change may be due to the dataset distribution. However, as shown in Fig.~\ref{pattern8}, PatternLDP cannot preserve shape information even given a large privacy budget $\epsilon=8$. In summary, the results highlight the efficiency of our optimization strategies for shape extraction.

\vspace{0em}
\subsection{Execution Time}
We evaluate the execution time at $\epsilon=4$. For the clustering task, we set the symbol size $t=6$ and the segment length $w=25$. While for the classification task, we set the symbol size $t=4$ and the segment length $w=10$. Note that we treat all the users' operations concurrently. And the results are averaged over 100 trials. We observed that PrivShape was slightly better than the baseline mechanism due to its more effective pruning strategy. Additionally, we found that PatternLDP primarily dedicated its time to  fitting the clustering or classification model.
\begin{table}[htbp]
	\vspace{-1em}
	\centering
	\caption{Execution time.}
	\begin{tabular}{|c|c|c|c|}
		\hline
		\textbf{Task} & \textbf{Baseline} & \textbf{PrivShape}&\textbf{PatternLDP} \\
		\hline
		\textbf{Clustering}& 1.88s & 1.69s & 9.98s \\ \hline
		\textbf{Classification} & 1.21s & 1.14s &133.82s \\
		\hline
	\end{tabular}
	
	\vspace{-1em}
\end{table}

\subsection{Parameters of SAX}
\begin{figure}[!htb]
	\vspace{-1em}
	\centering  
	\subfigure[$w=25$, varying $t$.]{
		\label{motionSymbol}
		\includegraphics[width=0.15\textwidth]{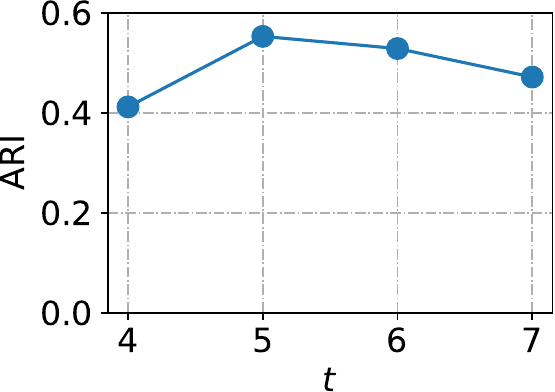}}
	\hspace{0.08\linewidth}
	\subfigure[$t=6$, varying $w$.]{
		\label{motionWindow}
		\includegraphics[width=0.15\textwidth]{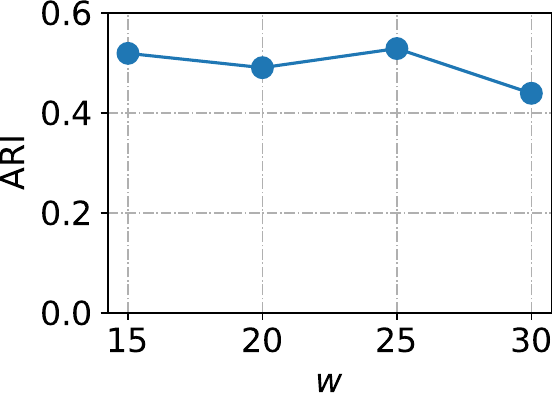}}
	\caption{Varying parameters for the Symbols dataset.}
	\label{motionpara}
\end{figure}

Our mechanisms have two parameters related to Compressive SAX: the symbol size $t$ and the segment length $w$. 
We set $\epsilon= 4$ and then vary the other two parameters.
Fig.~\ref{motionpara} demonstrates that the adjusted rand index (ARI) initially rises with an increase in symbol size $t$, but subsequently declines. With larger symbol sizes, more shape information is retained, resulting in improved clustering outcomes. However, larger symbol sizes also introduce fine-grained details, which can diminish the similarity. Similar trends are observed when the segment length varies in the results.

%
%
%

\begin{figure}[!htb]
	\vspace{-2em}
	\centering  
	\subfigure[$w=10$, varying $t$.]{
		\label{traceSymbol}
		\includegraphics[width=0.15\textwidth]{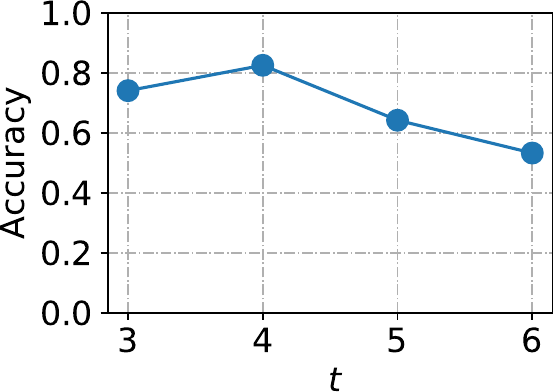}}
	\hspace{0.1\linewidth}
	\subfigure[$t=4$, varying $w$.]{
		\label{traceWindow}
		\includegraphics[width=0.15\textwidth]{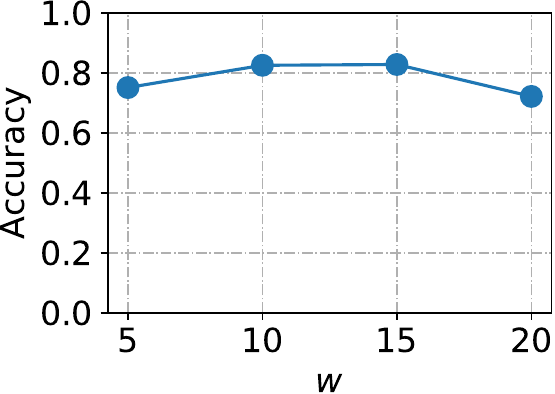}}
	\caption{Varying parameters for the Trace dataset.}
	\label{tracepara}
	\vspace{-1.5em}
\end{figure}

As  shown in Fig.~\ref{tracepara}, the  classification accuracy first increases as the symbol size $t$ increases but then decreases.  More symbols will extract more shape information, but the similarity of the shapes will decrease as too many fine-grained features are extracted. The results of varying the segment length exhibit a similar trend. When the segment length is small, more fine-grained features are captured, so similarity measuring is difficult. When the segment length is large, the extracted shapes lose too much information, leading to utility degradation.

\subsection{Impact of Distance Measure}
\begin{figure}[htbp]
	\vspace{-0.5em}
	\centering
	\includegraphics[width=0.35\textwidth]{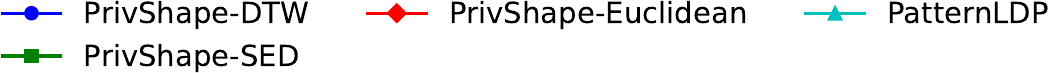}
	\vspace{0em}
	
	\subfigure[Clustering (Symbols)]{\includegraphics[width=0.16\textwidth]{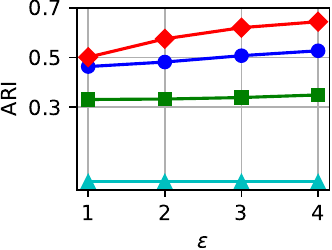}}
	\hfil
	\subfigure[Classification (Trace)]{\includegraphics[width=0.16\textwidth]{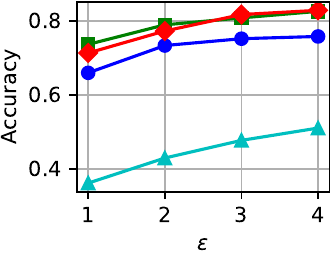}}
	
	\caption{Results from different distance metrics.}
	\label{diffdist}
	\vspace{-1em}
\end{figure}

In our experiment, we practically choose DTW~\cite{Kate2016} as the default distance measure for the clustering task and SED~\cite{Lin2007} for the classification task.  Here, we investigate the performance from three popular distance metrics for similarity measures between strings: DTW, SED, and Euclidean distance~\cite{Lin2007}. The results are shown in Fig.~\ref{diffdist}. As depicted in the figures, different distance measures yield different performance. However, all the results of PrivShape are superior to PatternLDP considering the practical privacy budgets $\epsilon\le4$. 

\subsection{Varying Time Series Length}
Intuitively, not only does the allocation of the privacy budget influence our mechanism's performance, but also the length of the time series. In this section, we investigate two conditions: the scenario where the shape remains constant despite variations in the time series, and the scenario where the shape changes as the time series varies. To evaluate the corresponding performance, we generate Trigonometric Wave as the dataset.  In the interest of space, we only show the result of classification task, and the task is for sine and cosine wave classification.  For PrivShape, we set the symbol size $t=4$ and segment length $w=10$. The parameters of PatternLDP are set the same as its original paper~\cite{Wang2020}.  Moreover, the privacy budget is set as $\epsilon=4$.

\subsubsection{Shape retains despite variations in the time series} 
\begin{figure}[htbp]
	\vspace{-1em}
	\centering
	\includegraphics[width=0.3\textwidth]{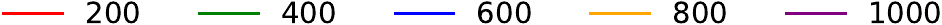}
	\vspace{0em}
	
	\subfigure[Sine Waves\label{sin_same}]{\includegraphics[width=0.2\textwidth]{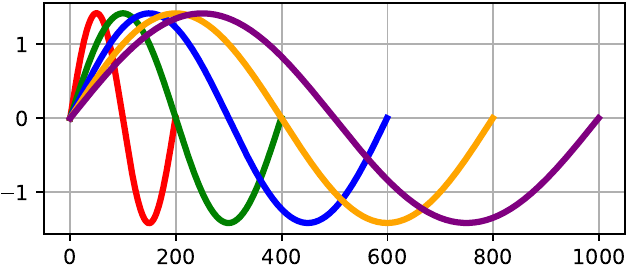}}
	\hfil
	\subfigure[Cosine Waves\label{cos_same}]{\includegraphics[width=0.2\textwidth]{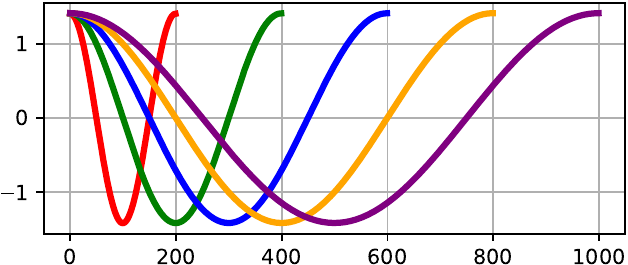}}
	
	\vspace{-1em}
	\subfigure[Classification on sine and cosine waves.]{\includegraphics[width=0.28\textwidth]{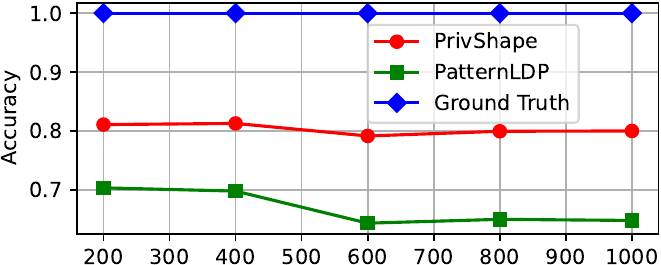}}
	\label{sameshape}
	\vspace{-0.5em}
	\caption{Varying time series length with the same shape.}
	\vspace{-0.5em}
\end{figure}

We generate sine and cosine values within a period with different lengths: 200, 400, 600, 800, and 1000, as shown in Figs.~\ref{sin_same} and~\ref{cos_same}.  To meet the requirement of PatternLDP, the generated time series are all z-score normalized. And the random forest classifier with the default settings~\cite{Pedregosa2011} is set as the ground truth. The classification results for sine and cosine waves are depicted in Fig. 16. The varying lengths only slightly influence our method due to the utilization of Compressive SAX among the same shape. However, as the length of the time series increases, the utility of PatternLDP is degraded.

\subsubsection{Shape changes as the time series varies} 
\begin{figure}[htbp]
	\centering
	\includegraphics[width=0.12\textwidth]{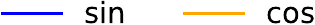}
	\vspace{0em}
	
	\subfigure[200]{\includegraphics[width=0.07\textwidth]{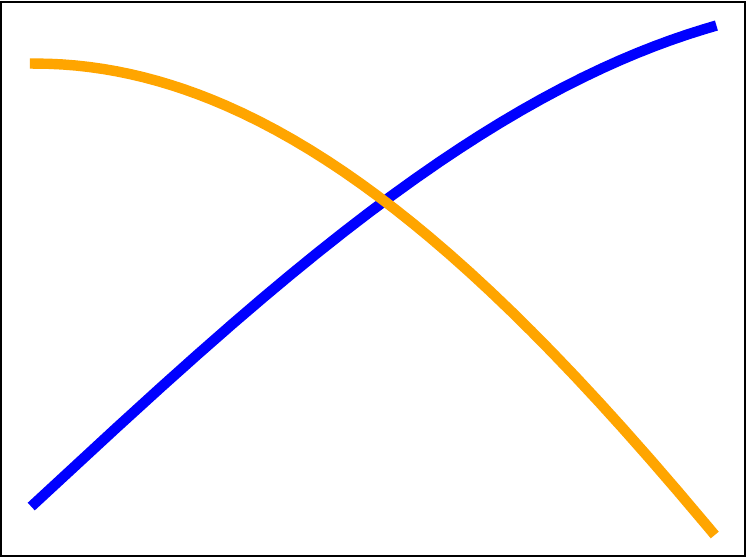}}\hspace{0.015\textwidth}
	\subfigure[400]{\includegraphics[width=0.07\textwidth]{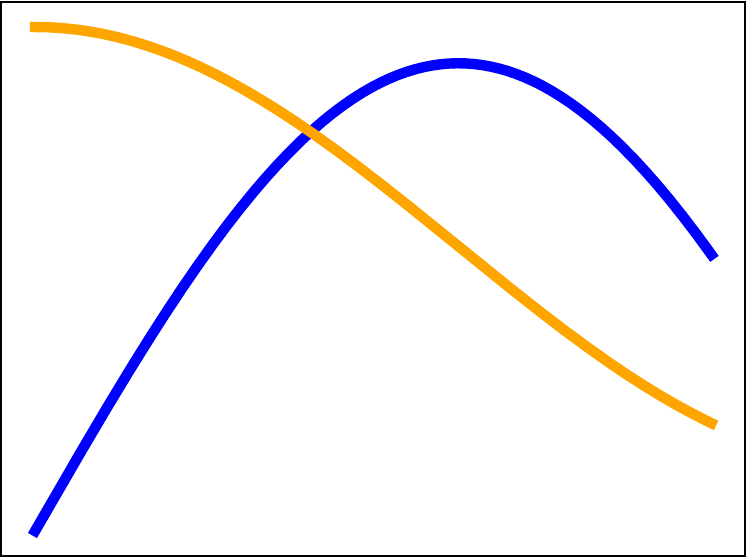}}\hspace{0.015\textwidth}
	\subfigure[600]{\includegraphics[width=0.07\textwidth]{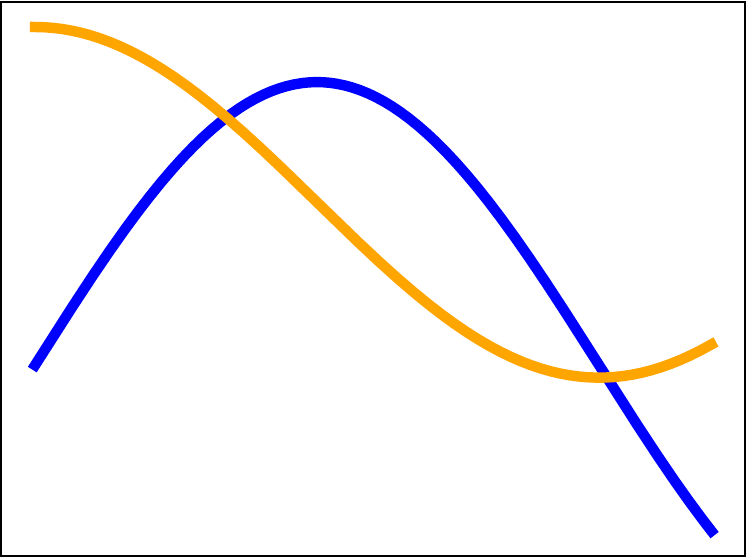}}\hspace{0.015\textwidth}
	\subfigure[800]{\includegraphics[width=0.07\textwidth]{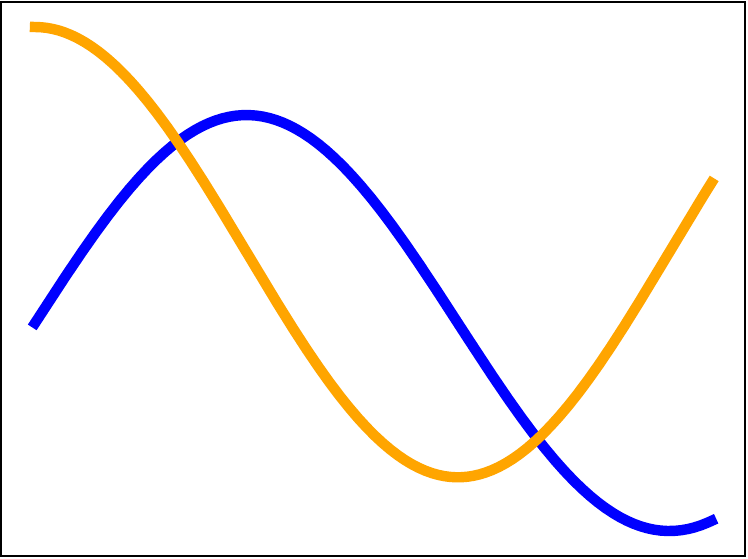}}\hspace{0.015\textwidth}
	\subfigure[1000]{\includegraphics[width=0.07\textwidth]{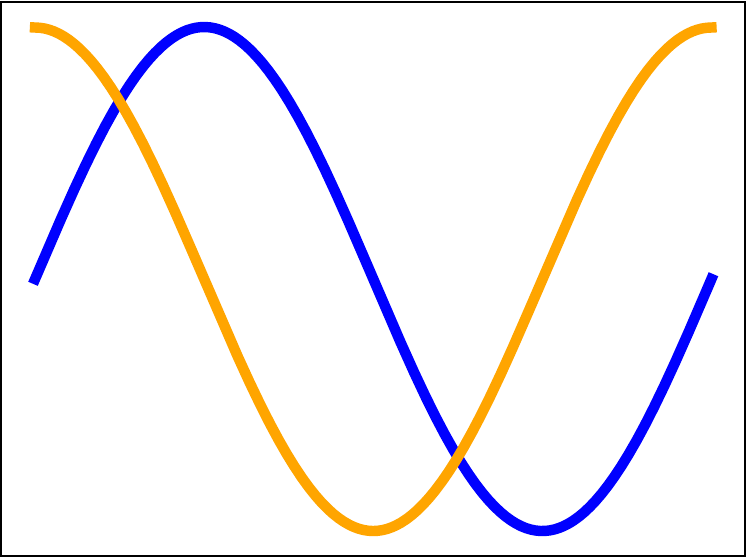}}
	
	\subfigure[Classification on sine and cosine waves.\label{diff_shape_result}]{\includegraphics[width=0.32\textwidth]{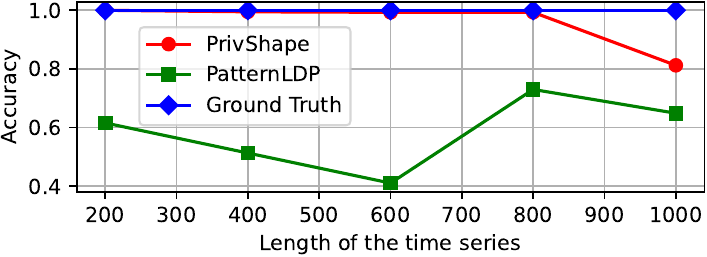}}
	\label{diffshape}
	\caption{Varying time series length with different shapes.}
	\vspace{-1.8em}
\end{figure}

We generate 1000-length time series consisting of sine and cosine values within a period, and select subsets of 200, 400, 600, 800, and 1000 data points, as shown in Figs.~17(a)-(e). And the data is also z-score normalized. The result can be found in Fig.~\ref{diff_shape_result}. We set the random forest classifier with the default settings~\cite{Pedregosa2011} as the ground truth. When the time series are similar in certain parts, PatternLDP cannot capture enough information to represent the shape, leading to significant fluctuations in the perturbation. However, the utility of PrivShape is still reasonable.

\subsection{Perturbation without SAX or Compressive SAX.}
\begin{figure}[htbp]
	\vspace{-0.5em}
	\centering
	\includegraphics[width=0.4\textwidth]{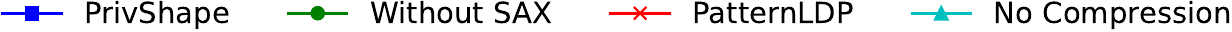}
	
	\vspace{0em}
	
	\subfigure[Without SAX \label{withoutsax}]{\includegraphics[width=0.18\textwidth]{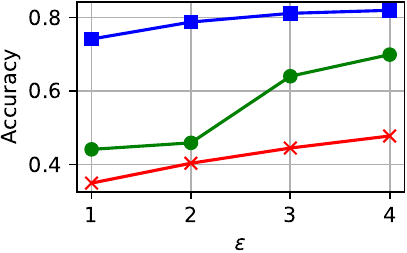}}
	\hfil
	\subfigure[No Compression \label{nocomp}]{\includegraphics[width=0.18\textwidth]{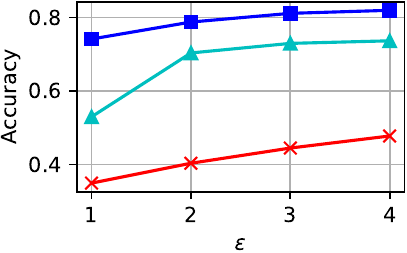}}
	\vspace{-0.5em}
	\caption{Results of ablation experiments.}
	\vspace{-0.5em}
\end{figure}

To show the advantage of our proposed mechanism, we conduct ablation experiments. In the interest of space, we only show the results of the classification task. Since PrivShape can only perturb discretized values, we discretize the values with 0.33 unit intervals starting from  0 and ending at 0.99 and -0.99, thus leading to eight segments on the y-axis. We set PatternLDP the same as in its original paper~\cite{Wang2020}. The results are shown in Fig.~\ref{withoutsax}. Since SAX captures the average value from a segment to mitigate the influence of noise and scaling, and the discretized allocation is closer to the original data distribution on the value-axis~\cite{Lin2007}, the pre-processed data by SAX contains more information about the essential shapes. Therefore, the utility of PrivShape without SAX is degraded,  but it is still better than that of PatterLDP. 

We also investigate the influence of the compression process after SAX (i.e., Compressive SAX), which mitigates the influence from time not warping by reducing the repeated symbols in the transformed SAX sequences. The parameters of PrivShape are set as the symbol size $t=4$ and the segment length $w=10.$ PatternLDP is set the same as in its original paper~\cite{Wang2020}. Although there will left more information without the compression process, the longer the sequence length, the fewer users each level of the trie will be allocated, leading to utility degradation. And the results shown in Fig.~\ref{nocomp} also validate the inference.

\section{Related Works}\label{relatedwork}
In this section, we review the literature on differential privacy and its application on time series release under LDP. In addition, we will discuss the work on frequent sequence mining under LDP.

\paragraph{Differential Privacy and Local Differential Privacy} Differential privacy in the centralized setting~\cite{Dwork2006} was initially introduced with a trusted third party and later extended to the local setting to accommodate broader scenarios~\cite{Ye2019, Ye2020a, Xue2023}. Many LDP mechanisms have been proposed for various tasks, such as frequency estimation~\cite{Wang2017, Wang2021} and mean estimation~\cite{Ye2019} in statistics collection. Recently, the research focus has been shifted to more complicated applications, such as itemset mining~\cite{Wang, Wang2018} and graph data mining~\cite{Ye2020a}.

\paragraph{Privacy-preserving Time Series Release under Differential Privacy}  Within the context of time series release, the mechanisms under DP can be categorized into three privacy levels: user-level privacy~\cite{Dwork2010}, event-level privacy~\cite{Dwork2010}, and $\omega$-event level privacy~\cite{Kellaris2014, Wang2016}. In 2010, Dwork et al.~\cite{Dwork2010} proposed the mechanisms for data stream release, with defining both user-level and event-level privacy. Subsequently, in 2014, Kellaris et al.~\cite{Kellaris2014} provided the definition for $\omega$-event level privacy. Due to the higher noise requirements, there are only several existing works focusing on user-level privacy under DP~\cite{Ahuja2023, Dong2023}. To improve the utility under user-level privacy, Ahuja et al. \cite{Ahuja2023} applied a sampling-based mechanism to decrease the elements sharing the privacy budget,  while Dong et al.~\cite{Dong2023} also deployed the advanced composition mechanism to enhance utility.

\paragraph{Privacy-preserving Frequent Sequence Mining under Local Differential Privacy} To the best of our knowledge, there are only three mechanisms for frequent sequence mining in the context of LDP. The first work, PrivTrie~\cite{Wang2018}, introduced an adaptive trie construction to mine the sequences with variable length and reuse  part of the users to enhance its utility. 
However, PrivTrie is not practically employed in a real-world scenario because it requires enormous communication resources to achieve user reuse~\cite{Wang2021}.  The Circular Chain Encoding (CCE)~\cite{Kim2020} aimed to estimate the frequency of new words by forming them from candidate $n$-grams and then verifying with pre-calculated fingerprints. 
Prefix Extending Method (PEM)~\cite{Wang2021} was proposed to estimate the frequency of categorical items with a large domain because  integer values can be transformed into certain bits.
Due to a small perturbation domain, PEM can extend multiple levels in a single round to allocate more users, thus enhancing its utility.

To the best of our knowledge, PrivShape is the first mechanism that addresses shape extraction from time series under user-level local differential privacy.

\vspace{0em}
\section{Conclusion}\label{conclusion}
In this paper, we propose a trie-based shape extraction mechanism, PrivShape, under user-level local differential privacy. We first employ Compressive SAX to reduce the processed elements. 
To enhance PrivShape's utility, we propose trie-expansion and two-level refinement strategies to reduce the trie's expansion domain and refine the frequency at the leaf nodes. Finally, we compare our mechanism PrivShape with the existing shape-retaining mechanism PatternLDP through conducting experiments on two benchmark datasets and one synthetic dataset. The experimental results show that our mechanism always outperforms and can obtain remarkable results with small privacy budgets even when $\epsilon\le2$. In general, we provide higher data utility with a stronger privacy guarantee. 

As for future work, we plan to extend this work to some practical applications, such as shapelets discovery, in order to address a broader range of real-world time series mining challenges.


\section*{Acknowledgment}
This work was supported by the National Natural Science Foundation of China (Grant No: 62372122, 92270123, 62072390 and 12371522), and the Research Grants Council, Hong Kong SAR, China (Grant No:  15225921, 15209922 and 15208923).


\end{document}